\documentclass[5p,times,twocolumn,preprint]{elsarticle}

\usepackage{hyperref}

\usepackage{amssymb}
\usepackage{amsmath}
\usepackage{mathtools}

\usepackage{subcaption}

\usepackage{soul}

\graphicspath{{figures/}}

\usepackage{xcolor}
\usepackage{listings}
\biboptions{sort&compress}

\newcounter{bla}

\journal{Computer Physics Communications}

\hypersetup{draft}

\newcommand{\MADX}{{MAD\textendash X}}

\begin{document}
\sloppy

\begin{frontmatter}



\title{BDSIM: An Accelerator Tracking Code with Particle-Matter Interactions}  


\author[rhul]{L.~J.~Nevay\corref{correspondingauthor}}
\ead{laurie.nevay@rhul.ac.uk}
\author[rhul]{J.~Snuverink}
\author[rhul]{A.~Abramov}
\author[ucl] {L.~C.~Deacon}
\author[rhul]{H.~Garcia-Morales}
\author[rhul]{H.~Lefebvre}
\author[rhul]{S.~M.~Gibson}
\author[rhul]{R.~Kwee-Hinzmann}
\author[rhul]{W.~Shields}
\author[rhul]{S.~Walker}
\author[rhul]{S.~T.~Boogert}

\address[rhul]{Royal Holloway, University of London, Egham, TW20 0EX, United Kingdom}
\address[ucl]{University College London, Gower Street, London, WC1E 6BT, United Kingdom}

\cortext[correspondingauthor]{Corresponding author}



\begin{abstract}
  Beam Delivery Simulation (BDSIM) is a program that simulates the passage of particles in a
  particle accelerator. It uses a suite of standard high energy physics codes
  (Geant4, ROOT and CLHEP) to create a computational model of a particle accelerator
  that combines accurate accelerator tracking routines with all of the physics processes of
  particles in Geant4. This unique combination permits radiation and detector background
  simulations in accelerators where both accurate tracking of all particles is required
  over long range or over many revolutions of a circular machine, as well as interaction
  with the material of the accelerator.
\end{abstract}


\end{frontmatter}



{\bf PROGRAM SUMMARY}

\begin{small}
\noindent
{\em Program Title: BDSIM}                                    \\
{\em Licensing provisions: GNU General Public License 3 (GPL)}\\
{\em Programming language: C++, flex, bison}                  \\
{\em External routines/libraries: Geant4, CLHEP, ROOT, gzstream, CMake} \\

{\em Nature of problem: Simulate energy deposition and charged particle detector background
  in a particle accelerator originating from beam loss where particles may pass both
  through the vacuum pipe with magnetic and electromagnetic fields,
  as well as through the material of the magnets and accelerator itself. Simulate the
  passage of particles both through an accelerator and the surrounding material such as air. Do so in
  a sufficiently flexible way that a variety of accelerator configurations can be easily
  simulated.}\\
{\em Solution method: Automatic creation of a 3D Geant4 model from an optical description
  of an accelerator using a library of generic 3D models that are user extendable. Accelerator
  tracking routines, the associated fields and coordinates transforms are provided for
  accurate magnetic field tracking.}\\
   \\


\end{small}

\section{Introduction}
\label{sec:introduction}

Particle accelerators are the primary tool to handle and study subatomic particles.
Originally developed to study particle physics, their applications are now widespread ranging from material
treatment in manufacturing to radio-nuclide production for medical imaging~\cite{Barbalat:260280}.
They are increasingly being used for the electromagnetic radiation they produce
in life-sciences to characterise biological samples~\cite{Arthur:2002ap, Balewski:2004iz, MAXIV}.

In every particle accelerator there are inevitable beam losses where some small
fraction of the particles transported are not contained, whether by design or
otherwise. These may be lost due to 
the initial momentum distribution of the source particles and the finite extent
of the accelerator components and their fields. Non-linear or time-varying fields may
also lead to a `dynamic aperture' that is typically smaller than the physical aperture
that results in further losses. Any losses, no matter how small, can
lead to detrimental effects such as heat load in components, experimental backgrounds,
reduced machine availability and long term radioactivation and damage.

In the case of high energy accelerators, even minute losses can lead to problematic
radiation or heat loads. As the energy per particle increases, so does the distance the particle
can penetrate in material. In the case of the very highest energy accelerators such as
the Large Hadron Collider (LHC) at CERN~\cite{LHCDesignReport} with 6.5\,TeV protons,
particles may penetrate tens of metres of rock or concrete. Particles may also quasi-elastically
scatter with low momentum transfer on the edge of part of the accelerator such as a
collimator and continue to a distant part of the accelerator. Therefore, the locations
of energy deposits and particle losses are not always
directly correlated with the scattering point.

Many modern accelerators including the LHC make use of cryogenic equipment such as
superconducting magnets or radio frequency cavities to achieve the required high magnetic
or electro-magnetic fields~\cite{Lebrun_2017}. Such cryogenic equipment must be kept
below a certain temperature limit as well as within a narrow range to remain superconducting
otherwise they will quench. Accelerators that make use of cryogenic equipment will
have systems to safely cope with a quench, however, it will take many hours to
recover to an operating state, primarily due to the inefficiency of cooling at
cryogenic temperatures as well as low thermal conductivity. Such events reduce
the accelerator availability and can limit its utility, such as integrated
luminosity and data collected. For these reasons, losses must be precisely predicted
as well as the location of the subsequent energy deposition.

When a particle beam is stored for minutes to hours in a storage ring collider, various
effects lead to the formation of a beam halo\textemdash{}particles that follow the main beam but
with a large amplitude~\cite{Valentino:2013}. Beam halo must be continuously removed
to avoid increased energy deposition and to protect both the accelerator and any
detector close to the beam. In high energy accelerators it is often not possible
to remove the halo with one stage collimation as the required length of
material is prohibitively long to be put in one location. Single and multi-stage
collimation systems require a simulation involving the interaction with matter
to accurately predict their efficiency as well as any leakage of collimation
products~\cite{Bruce:2014}.

A further consideration is the interface between the accelerator and a possible
detector or experiment, commonly referred to as the Machine Detector Interface (MDI).
The beam size is often strongly focussed to create a small transverse beam size
at the centre of the target area to increase the collision rate
between the two crossing beams. This manipulation can lead to increased losses around
a collision point and therefore background
radiation that may penetrate the detector\textemdash`non-collision background'. Particles
in the beam may also collide with any residual gas molecules in the vacuum of the accelerator
beam pipe leading to similar background through the machine and into the detector. 
Such backgrounds may give the appearance
of potential new physics if not accurately accounted for. Although the direction or
timing of such signals can be used to help discriminate against genuine collisions,
overall the background should be minimised as much as possible to avoid degradation
in the ability of the detector to correctly identify the collision events.

In each of the aforementioned scenarios, a simulation is required that includes both
tracking of particles through magnetic and electro-magnetic fields of accelerators as well as
the interaction with the material of the accelerator including the production and
transport of secondary radiation. Simulations of these two features have historically
been separate simulations each with their own specialised tools.

To predict the losses throughout a machine, a trivial estimate can be made by comparing
the aperture to nominal beam size throughout the machine. However, the nominal beam size
is typically derived from the linear lattice functions and does not account for the variation
in particle momentum, any nonlinear fields, nor a non-Gaussian beam. Therefore, to accurately
predict the
losses, a particle tracking simulation is performed. A particle distribution is
sampled and a Monte Carlo simulation performed by calculating the individual trajectories
of particles until a termination condition is reached. Such a condition may be a single
passage through a model or a certain number of revolutions of a circular accelerator.
If the aperture is included in the simulation, particle tracking may be stopped when the
particle position exceeds that of the aperture boundary.

A common tool for accelerator design is \MADX{}~\cite{MADX-website}. This provides the ability
to define a sequence of magnets, calculate the linear lattice functions as well as
track particles using the Polymorphic Tracking Code (PTC)~\cite{Schmidt:2002vp}.
\MADX{} is commonly used to prepare an input model for the
SixTrack tracking code~\cite{Schmidt:1994kc} for long-term symplectic tracking and
dynamic aperture studies for the LHC. In
both the case of \MADX{} PTC and SixTrack, the particles are tracked throughout the complete
model irrespective of aperture information. To estimate losses, the output trajectories
can then be filtered by an independent program to apply the aperture model and terminate
the trajectories at the appropriate point. The termination
points of all the trajectories can then be collated to form a loss map~\cite{Bruce:2014}.
However, this is a specialised workflow rather than a publicly available tool. Whilst
such a strategy is a demonstrably successful one, the simulation
stops at the point where the particle touches the aperture.

However, upon impact with material high energy particles will disintegrate or
create particles, creating large
amounts of radiation on a length scale that increases logarithmically with the
energy of the incident particle. For a given kinetic energy of a
particle travelling in a material, a stopping distance can be calculated and it could
be assumed that although the impact is not simulated, any subsequent radiation would
occur on this length scale.  However, as the particles are relativistic, they
impact and scatter at very low angles and so it's possible for a particle to
re-enter the vacuum pipe. In this case the particle
may travel quite some distance before again impacting the aperture. In the case
of nuclei, fragmentation may occur producing nuclear fragments or protons with
a momentum inside the acceptance of the accelerator and therefore travel a great
distance. It is therefore crucial to simulate the interaction of losses with
the accelerator as well as their subsequent propagation and secondary radiation
to make an accurate prediction. For charged particle background in a detector
it is also crucial to simulate the interaction with the accelerator as the
background is predominantly composed of secondary particles produced by the accelerator.

Two further unique scenarios that require combined accelerator and particle matter
interactions are muon backgrounds for detectors and secondary
beam production and transport. In the first case, with the impact of high energy
particles, secondary muons can be produced. These can penetrate large distances
through material and may pass through the magnets and vacuum pipe including their
magnetic fields. Being so penetrating and charged, they often contribute strongly
to detector background. Secondly, the production of exotic secondary beams, is
typically achieved by colliding a beam with a fixed target. To accurately understand
the spectrum of transported particles it is vital to simulate the particle-matter
interactions as well as the transportation through electromagnetic fields.

Simulations that handle the interaction with matter are commonly made to predict
particle detector response and
accuracy. A 3D model with material specification is required as well as a
library of particles and physics processes. If a magnetic or electric field
is present, support for this is also required. Geant4~\cite{GEANT4:2003} and
FLUKA~\cite{FLUKA:2015} are two software packages that provide the capability
to simulate the passage of particles in matter. Geant4 provides an open source
C++ class library where the user must write their own program to construct
the geometry and run the simulation. The open source nature allows users to
introduce their own functionality through many interfaces. Geant4 also includes
an easy-to-use visualisation engine and interactive interpreter to drive the simulation.
FLUKA is a closed-source Fortran code where the
user describes their model through input text files. In both cases,
significant effort is
required to describe the geometry and materials particular to a given
experiment or accelerator to be simulated. Furthermore, the user must supply
a numerical or functional field map for each volume they require to have a magnetic or
electric field.

Numerical integration is used to calculate the particle motion in an arbitrary field,
which while flexible, can suffer from the accrual of small numerical errors that can eventually lead to
gross inaccuracies if used repeatedly. Limiting these effects by permitting only
small steps in the field may make the simulation prohibitively computationally
expensive as each high energy `primary' particle may lead to thousands of
`secondary' particles that all must be tracked through the field. For the
purpose of motion in an accelerator, numerical integration is often not suitable as
it is not sufficiently accurate after the many steps required through the
large number of different magnetic fields, hence the use of dedicated
accelerator tracking codes. However, accelerator tracking codes do not provide the
physical processes or the 3D geometry required for the required simulation.

In the past, attempts have been made to solve this problem. One such historical example is
TURTLE~\cite{Brown:1974ns}, which was originally an optical transport code with the inclusion
of some in-flight decay physics processes that was subsequently modified to
include duplicated parts of REVMOC~\cite{Kost:1983fy}, a Monte Carlo tool for particle matter interaction,
to include a limited set of physics processes most relevant to scattering in
collimators. However, this legacy Fortran code was developed by many groups
around the world resulting in several versions. With the availability of more
modern programming languages and parser generators for input syntax, legacy
Fortran input cards are not easily understood nor used. Geant4 includes a few
example applications that demonstrate a beam line simulation. However, these
do not include accelerator tracking routines and use only numerical integration
for particle motion; have typically less than 10 components; and are hard-coded
to that beam line experiment in C++.

Beam Delivery Simulation (BDSIM)~\cite{BDSIM-website,BDSIM-manual} is a program that
solves this general problem of mixed accelerator and particle matter interaction
simulations by creating a 3D model using the Geant4 library with
the addition of accelerator tracking routines. Geant4 was chosen as it is
open source and so permits the extension of tracking routines as well as being
written in a more modern flexible language.

The code described in this letter, BDSIM, is in fact a completely rewritten
version of BDSIM described in~\cite{Agapov:2009zz}, which commenced in 2013. The
core implementation has been entirely revised and validated and whilst the premise
is the same, they cannot be considered equivalent. A comparable code is
G4beamline~\cite{g4beamline-website},
which allows creation of a Geant4 model of beam lines. However, this was first
started after the original version of BDSIM. G4beamline has more simplistic geometry
in comparison to BDSIM with only simple cylinders for each magnet, and also cannot
simulate circular machines. BDSIM is also set apart with its advanced scalable output
and per-event analysis described in later sections permitting complicated analyses.

Accelerators are typically constructed
with as few classes of magnet as possible and feature repetitive patterns
of sets of magnets. Whilst the aperture of the vacuum pipe may vary in
size, most designs fall into a small set of cross sections. BDSIM provides a library of
scalable and customisable 3D components that provide the most commonly used magnets
and apertures for an accelerator.

BDSIM constructs a 3D model using this
library from an optical description of an accelerator, i.e. one that
describes the length, type and strength of each magnet in a sequence.
Along with each 3D model of the different types of magnets, appropriate
fields are provided that are calculated from the rigidity-normalised
strength parameters mostly commonly used to specify accelerator magnets.

With BDSIM, the user may progress from a generic model to a more specific
one by adding externally provided geometry for particular elements, or by
placing such geometry beside the accelerator in the model.
Users can overlay their own field map on top of parts or all of components
and choose between provided numerical interpolators for the discrete map.
A human-readable input syntax is used so the user may provide input text
files to describe the model and need not write code nor compile it.

BDSIM provides a unique simulation capability that can also be accessed
in a very short timescale from an optical accelerator description. The
distinctive capabilities allow both energy deposition throughout an accelerator
to be simulated as well as interfaces between accelerators and detectors. The
implementation and a worked example highlighting the features are described
in the following sections.

\section{Implementation}
\label{sec:implementation}

Geant4 is a C++ class library that provides no standard program the developer or user
can run. A developer must write their own C++ program to instantiate classes
representing geometrical shapes, materials, placements of shapes in space
as well as physics processes and the Geant4 kernel. As C++ is a compiled
language, this would generally make any Geant4 model fixed in design. However, to
simulate any accelerator, a more dynamic setup is
required.

BDSIM uses human readable text input files with a syntax called \emph{GMAD}. The GMAD
syntax (Geant4 + MAD) is designed to be as similar as possible to that of MAD8 and \MADX{} that are
common tools for accelerator design and therefore it will be immediately familiar
to a large number of users in the accelerator community. This is significantly
less labour intensive than writing and compiling C++ code.

BDSIM uses GNU Flex~\cite{FLEX-website} and GNU Bison~\cite{BISON-website}
to interpret the input text files and prepare the necessary
C++ structures for BDSIM to create a Geant4 model. Use of a parser ensures
strict compliance with the language and sensible user feedback to prevent
unintended input or mistakes. The parser is easily extended by
the developer allowing the possible introduction of new features in future. The
most minimal input includes

\begin{enumerate}
\item At least one beam line element.
\item A sequence (`line') of at least one element.
\item Declaration of which sequence to build.
\item The particle species.
\item The particle total energy.
\end{enumerate}

\noindent and would be written as

\begin{lstlisting}
d1: drift, l=1*m;
l1: line=(d1);
use, l1;
beam, particle="e-",
       energy=10*GeV;
\end{lstlisting}

This model defines a single beam pipe with no fields and an electron beam of
10\,GeV total energy. Additional options and sets of physics processes may
also be specified. After parsing the
input text files, the Geant4 model is constructed by instantiating various
construction classes that are registered with the Geant4 kernel class \texttt{G4RunManager}.
The various aspects of the model construction are described in
subsections~\ref{ssec:geometryconstruction}\textendash\ref{ssec:control}.

\subsection{Geometry Construction}
\label{ssec:geometryconstruction}

The model is built from a sequence of unique elements that may appear multiple times
in a varied order. As there are 26 different elements defined in BDSIM including 12
types of magnets with 8 different styles that can be combined with any one of
8 aperture cross sections, there is a large number of possible geometry combinations.
It would be impractical to have one C++ class for each combination and it would not
be trivial to extend the code to include new aperture cross sections or magnet styles.
BDSIM is therefore designed in such a way that independent pieces of geometry can be constructed
and then placed safely either alongside each other or in a hierarchy. This allows beam pipes
and magnets to be constructed independently and assembled. Furthermore, it makes extension
to include new aperture shapes or magnets trivial. Independent factory pattern classes
for magnet yokes and beam pipes allow any combination of
aperture and magnet style to be created. Therefore, only one class is required for a magnet
that uses the factories to create the yoke and beam pipe in the desired combination.

Geometry in Geant4 is constructed using the constructive solid geometry (CSG) technique,
where primitive shapes (e.g. a cube, sphere, cylinder, etc.) can be combined through
Boolean operations (union, subtraction, intersection) to create more complex shapes.
The program developer instantiates classes with the relevant parameters and places
them together inside a parent volume. With the Geant4 geometry system, it is entirely
possible to construct a nonphysical
geometry where a volume overlaps with other volumes at the same hierarchy level, or protrudes
outside the containing volume. When tracking a particle through the geometry, only other
volumes at the current level of the geometry hierarchy are searched for where the track will
enter them. If, through a combination of bad placement and shape parameters, the particle
wrongfully appears already inside an overlapping volume, the volume will not be detected as
the particle has not \emph{entered} it. Such errors are only highlighted to the user if they
purposefully scan the geometry for errors or worse, during a simulation when the
tracking routines fail to navigate the geometry hierarchy correctly resulting in
material being skipped or the particle jumping through the geometry. This may proceed
without any error and simply produce an incorrect result. When creating a Geant4 model,
we must therefore be careful to ensure no overlaps exist, which is especially necessary
with a parameterised model defined by user input. BDSIM ensures that any user input
using BDSIM's standard components will result in a safe Geant4 model without overlaps,
including any combination of the available geometries (e.g. aperture shapes and magnets).

Once one of the Geant4 CSG primitive classes is instantiated, it is not
possible to know its extent without querying a tracking point as to whether it
lies inside or outside the volume. To circumvent this as well as prevent
construction of invalid geometry, BDSIM records the
asymmetric extents in three dimensions of every piece of geometry created.
Whilst the cuboid denoted by these extents does not represent the surface of
the volume, it is sufficient for ensuring that no overlaps will occur. Any piece of
geometry in BDSIM is therefore represented by the base class \texttt{BDSGeometryComponent}
that handles the extents.

Furthermore, to correctly navigate the geometry hierarchy, Geant4 must be able to numerically
determine whether a point in 3D Cartesian coordinates lies inside or outside of a volume.
Therefore, two volumes placed adjacent to each other at the same level in the
geometry hierarchy must have a non-zero space between them. Geant4 defines a
geometry tolerance that is the minimum resolvable distance in the geometry
and therefore the tolerance when estimating the intersection with a surface
of a trajectory. The tolerance is set explicitly in BDSIM to 1\,nm and
this is defined as a constant throughout the code called `length safety' that
is used to pad all geometry hierarchy.

The geometry is constructed by the \linebreak \texttt{BDSDetectorConstruction} class that uses
a component factory (\texttt{BDSComponentFactory}) to create the individual components
required. A component registry is used to reuse previously constructed components
saving a considerable amount of memory for large models. Components whose field
is time dependent and therefore depends on the position in the beam line are
created uniquely to ensure correct tracking. Each component is appended
to an instance of \texttt{BDSBeamline}, which interrogates each element and prepares
the 3D transformations (rotation and translation) required to place that element on the
end of the beam line in 3D Cartesian coordinates. When the construction of the beam line
elements is complete, they are placed in a single container `world' volume.
Between construction and placement, the physical extent of the beam line is
determined and these are used to dynamically construct a world volume of the
appropriate size for the model.

Some elements may make use of geometry provided in external files. Such geometry
is constructed again through a factory interface with
a different loader for each format supported. The primary format is
Geometry Description Markup Language (GDML)~\cite{GDML:2006}, which is the geometry
persistency format of Geant4.
External geometry can either be placed in sequence in the beam line, wrapped
around a beam pipe as part of a magnet or irrespective of the beam line in
the world volume at an arbitrary location.

For increased physical accuracy, BDSIM is capable of building a tunnel around
the beam line. Accelerators are typically placed in a shielded environment to contain
any radiation and high energy accelerators are typically placed in tunnels underground. The tunnel
built by BDSIM can be customised by the user to be one of 5 different cross sections
and offset with respect to the beam line. The algorithm used to generate the tunnel geometry
is capable of following complex beam lines unique to the user input model. The tunnel
is highly suited to prevent
cross-talk between parts of the model in a circular model that would otherwise be an artefact
of simplified geometry of the simulation. Options allow particles to be removed from
the simulation when they impact the tunnel for computational efficiency.

Due to the Geant4 interface, the fields for each element are not constructed
at the same point as the geometry. Geant4 requires all fields to be constructed
and attached to volumes at one point in the program. When the geometry
for an element
is constructed, a field \emph{recipe} and logical volume to attach it to are
registered to a field factory. The factory is then used by the Geant4 interface
to construct and attach all fields at once.

\subsection{Coordinate Systems \& Parallel Worlds}
\label{ssec:parallelworlds}

The majority of accelerator magnetic fields as well as externally provided field
maps are usually defined with respect to the local coordinate frame of the element they are attached to.
Similarly, accelerator-specific numerical integration algorithms for calculating a particle
trajectory through an element are typically defined in a curvilinear coordinate frame
that follows the trajectory of a particle with no transverse position or momentum and
with the design energy through that element\textemdash{}the Frenet-Serret coordinate
system. However, contrary
to this, Geant4 uses the 3D Cartesian coordinates described by the outermost volume
in the geometry, i.e. the world volume. BDSIM provides the necessary coordinate
transforms between these systems to permit the use of accelerator tracking routines
and field maps attached to individual elements.

\begin{figure}
  \centering
  \includegraphics[width=9cm]{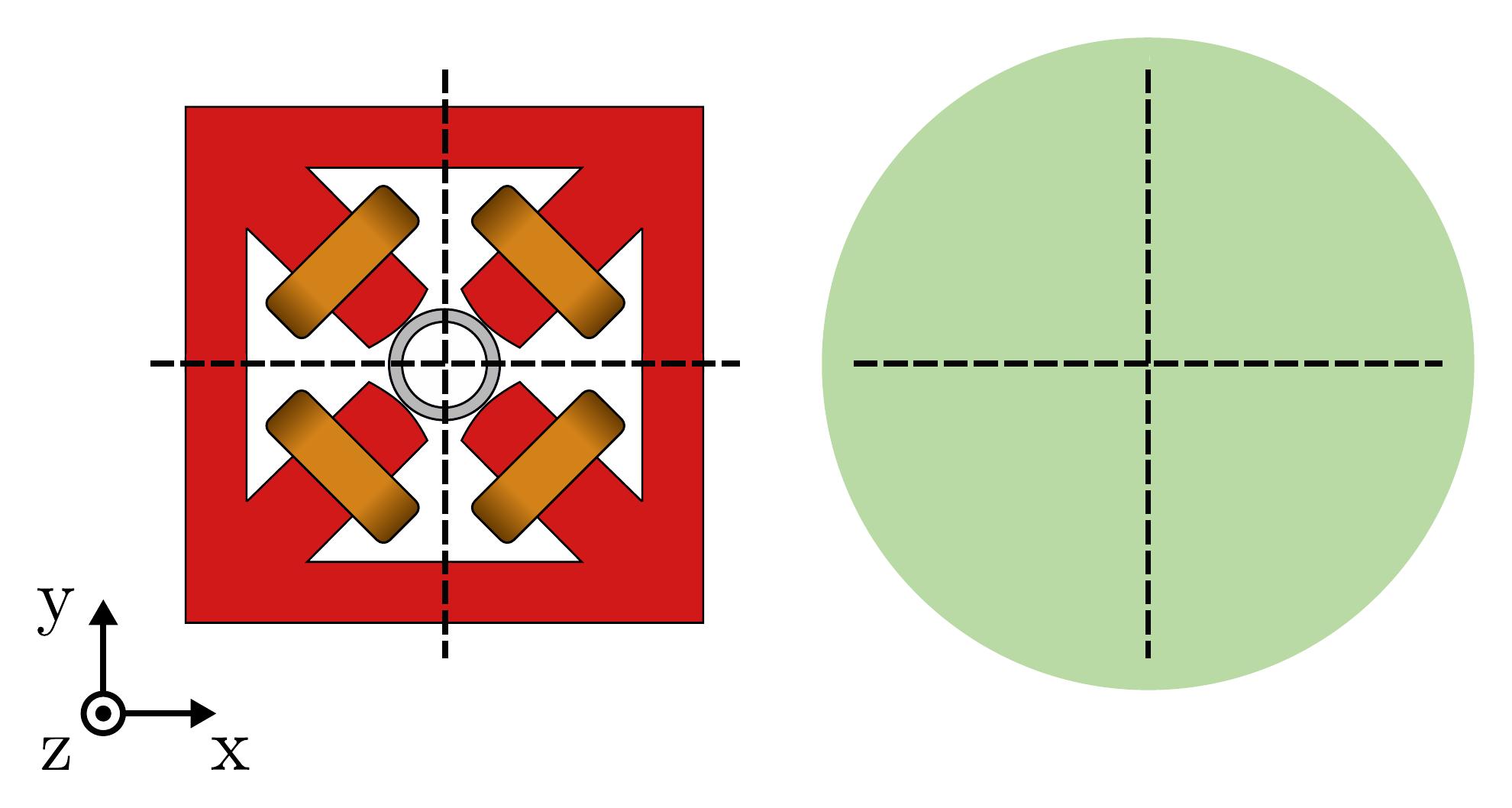}
  \caption{\label{fig:curvilinearcylinder}Comparison view of quadrupole in the mass
    world and the accompanying cylinder used for curvilinear coordinate frame transforms
    in a parallel world. The curvilinear cylinder has been reduced in size for comparison, but
    is usually several times bigger to overlap all of the beam line.}
\end{figure}

For a given global Cartesian position, Geant4 can provide the transform
from the volume that point lies within to the world
volume and vice-versa irrespective of the depth of the
geometry hierarchy. A transform for $N$ levels higher in the geometry hierarchy
can also be requested. As the depth of the geometry hierarchy may vary from
component to component, this facility cannot be used generally. The local coordinate frame
of any given volume is also not necessarily the required curvilinear frame.
To provide the correct transforms into the curvilinear frame, BDSIM constructs
a separate 3D model (a `parallel world' in Geant4 terminology) with different
geometry than that of the beam line. In this parallel world, a simple cylinder
of the same length as the accelerator component is placed at the same position
as shown in Figure~\ref{fig:curvilinearcylinder}.

Any point in the world can then be queried in the parallel world and the transform used from
the volume found at that location to the world volume. This will be a transform
from the world volume to the local coordinate system of the cylinder whose axis
is degenerate with the curvilinear system required. In the case of a component
that bends the beam line, many small straight cylinders with angled faces are
constructed. This is not exactly the same as the true curvilinear frame, so a
dedicated transform is provided.

\begin{figure}
  \centering
  \includegraphics[width=8cm]{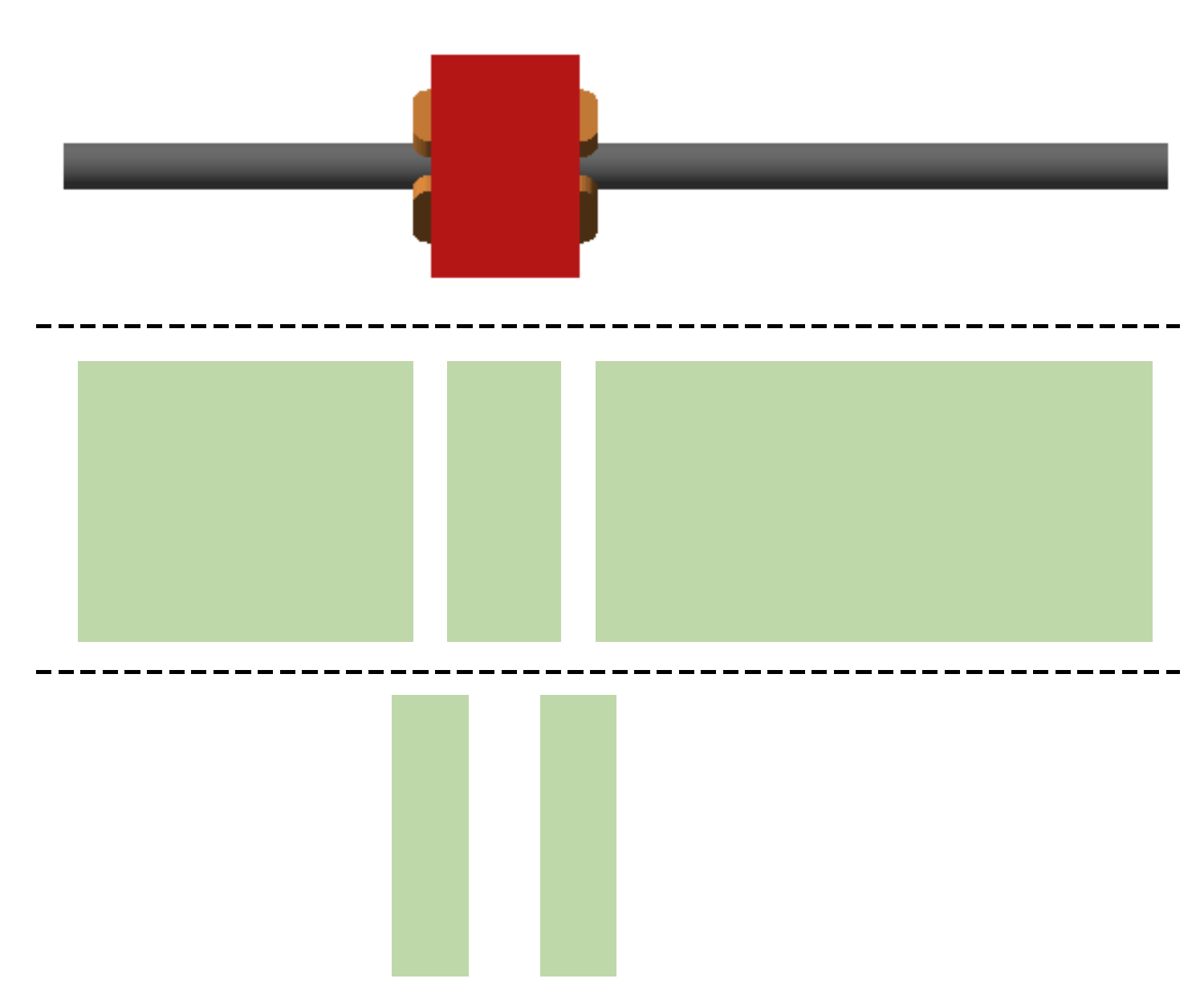}
  \caption{\label{fig:bridgingworld} Side view showing the geometry in three different
    worlds separated by a dashed line. The top-most is the mass world with a beam pipe,
    quadrupole (red), and beam pipe. The middle world shows three curvilinear cylinders
    (green) with non-zero gaps between each. These are used for the curvilinear transforms.
    At the bottom, the third world shows curvilinear bridging sections (green) designed to
    overlap the above curvilinear cylinders but also have a non-zero gap between them. The
    bridging cylinder length has been exaggerated for visualisation purposes.}
\end{figure}

As already discussed, all Geant4 geometry must have a numerically resolvable
gap between adjacent solids and so each parallel world cylinder is placed
with a small gap between it and the next one in the beam line. However, if
a point is queried in this gap, an incorrect transform will be found. This occurs
regularly as the particle enters or exits a volume. The gap between volumes is kept
to a minimum to avoid integrating the field over a shorter length. To overcome
this problem, a third parallel world is built with bridging cylinders. While searching
the curvilinear world for a volume, the world volume itself is found, then the bridging
world is subsequently searched. This ensures a continuous coordinate system irrespective of
the limits of the geometry system. A quadrupole with curvilinear cylinders and bridging
cylinders are shown in Figure~\ref{fig:bridgingworld}.

\subsection{Fields}
\label{ssec:fields}

Along with the geometry of each beam line element, the required fields are
provided. The field strengths are calculated from the rigidity-normalised
strength parameters most commonly used to specify accelerator magnets
and used in accelerator modelling tools such as \MADX{}. For example, in
the case of a quadrupole, the normalised strength $k_1$ is typically used,
which is defined by

\begin{equation}
  \label{eq:k1}
  k_1 = \frac{1}{B\rho} \frac{\partial B_{y}}{\partial x}
\end{equation}

\noindent where $B\rho$ is the magnetic rigidity of the particle and $B_y$
is the vertical component of the magnetic field. The magnetic rigidity is given by

\begin{equation}
  B\rho = \frac{p}{e}
\end{equation}

\noindent where $p$ is the momentum of the particle and $e$ is the elementary
charge. $B\rho$ is calculated for the beam particle the accelerator is designed
for and this is used to calculate the magnetic field gradient for a given
$k_1$. The gradient is evaluated by rearranging Equation~\ref{eq:k1} and then
evaluating it at the coordinates of the particle with

\begin{equation}
  \begin{aligned}
    B_x &= x \, \frac{\partial B_{y}}{\partial x}, \\
    B_y &= y \, \frac{\partial B_{y}}{\partial x}, \\
    B_z &= 0
  \end{aligned}
\end{equation}

This is valid close to the axis of the magnet as it neglects any description
of the field close to the magnet poles. The \emph{pure} field like this is
used for the vacuum and beam pipe volumes. For the magnet yoke and air in between
the poles, a different description is used. Here, a 2D field, invariant along the
length of the magnet, is used that is the sum of point current sources located at
the point half way between each pole, where the coils would be. The location of
each wire is given by

\begin{equation}
   \mathbf{c}_i =
   \begin{bmatrix}
   x \\
   y \\
   \end{bmatrix}_i
   =
   \begin{bmatrix}
   0  \\
   r_{\mathrm{pole tip}} \\
   \end{bmatrix}
   \begin{bmatrix}
   \cos \theta_i & - \sin \theta_i \\
   \sin \theta_i & \cos \theta_i   \\
   \end{bmatrix}
\end{equation}

\begin{equation}
  \theta_i = \left \{ \frac{i\,2\pi}{n_{\mathrm{poles}}} \right \} \quad \mathrm{for} \quad i = \{0 \ldots n_{\mathrm{poles}} \}
\end{equation}

\noindent where $r_{\mathrm{poletip}}$ is the radius at the pole tip and $\theta_i$ the angle
in polar coordinates of the $i^{th}$ wire current source. The magnetic field value $\mathbf{B}$
as a function of position $\mathbf{r} = (x,y)$ is

\begin{equation}
  \mathbf{B}(\mathbf{r}) = \sum_{i = 1}^{n_{\mathrm{poles}}} (-1)^{i} \, \frac{(\mathbf{r} - \mathbf{c}_i)_{\perp}}{\|\mathbf{r} - \mathbf{c}_i\|}
\end{equation}

\begin{figure}
  \centering
  \includegraphics[width=9cm]{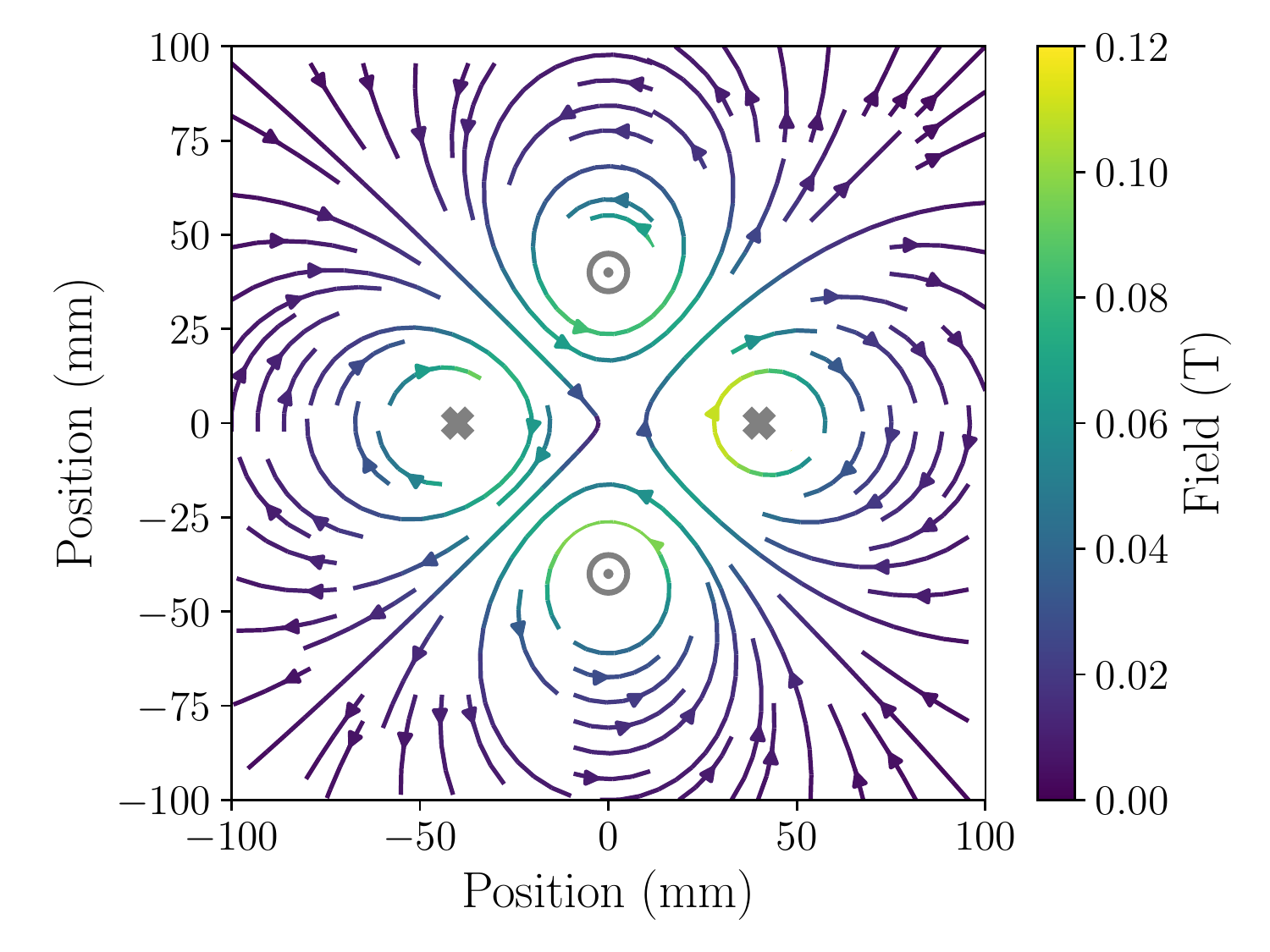}
  \caption{\label{fig:multipolefield}Field for a quadrupole yoke with a pole tip radius of
    40\,mm as a function of local transverse coordinates. The direction of the current
    is shown by the grey cross and points at each current source location.}
\end{figure}

The field is normalised to the pure field at the pole tip radius. This approximate
field neglects the permeability of the yoke material, but is sufficient for the
purpose as only secondary particles are expected to be transported outside the beam
pipe. Should the user wish, they may overlay their own field map for a more accurate
description. An example field of this form for a quadrupole is shown in
Figure~\ref{fig:multipolefield}.

\subsection{Physics Processes}
\label{ssec:physicsconstruction}

Geant4 provides a large library of physics processes. A Geant4 application must
instantiate the required particles and then instantiate the required physics
process classes and attach them to the applicable particles. Geant4 includes
a more convenient general set of `physics lists' that provide commonly used sets
of physics processes applicable
to many particles for a given application and energy range. These modular
lists can be selected by the user in BDSIM and it is left to them to select the most
relevant physics processes for the simulation. Using all of the physics
processes by default would result in an extremely computationally expensive
simulation. Additionally, there may be more than one physics model relevant
that the user may wish to choose from. BDSIM provides simple names that
map to the most common physics lists constructors in Geant4 as described in
Table~\ref{tab:physicslists}.

\begin{table}
  \centering
  \caption{\label{tab:physicslists} Examples of name mapping of Geant4 physics
    lists in BDSIM.}
  \begin{tabular}{l l}
    \hline \hline
    BDSIM Name         & Geant4 Class                      \\ \hline
    em                 & G4EmStandardPhysics               \\ 
    em\_livermore      & G4EmLivermorePhysics              \\ 
    decay              & G4DecayPhysics                    \\ 
    ftfp\_bert         & G4HadronPhysicsFTFP\_BERT         \\ 
    ftfp\_bert\_hp     & G4HadronPhysicsFTFP\_BERT\_HP     \\ 
    hadronic\_elasitc  & G4HadronElasticPhysics            \\ 
    ion                & G4IonPhysics                      \\ 
    shielding          & G4ShieldingPhysics                \\ 
    stopping           & G4StoppingPhysics                 \\ 
    synch\_rad         & G4SynchrotronRadiation            \\ \hline \hline
  \end{tabular}
\end{table}

If no processes are selected, only tracking in magnetic fields will be present,
i.e. particles will pass unimpeded through matter. Such process-less models are
useful for validating particle tracking and beam distributions as well as
comparisons to other codes that lack such interactions.

\subsection{Run and Event}
\label{ssec:runandevent}

A BDSIM simulation progresses at two levels; a \emph{Run} and an \emph{Event}. An event
is the smallest unit of simulation where one initial particle or set of particles
is tracked through the model. All information may be collected on an event level
basis and all events are entirely independent of each other. A run is a unit of
simulation containing $N$ events where the geometry and physics processes do not
change.

BDSIM can be run in two ways; interactively with a visualiser; or in batch mode
without visualisation. The latter is considerably faster and used for large scale
simulation, whereas interactive visualisation is typically used in the preliminary
stages to verify the model and typical outcome of an event. In the case of batch mode,
BDSIM performs one run with the desired number of events. Interactively, the user
may issue the following command on the interactive terminal

\begin{lstlisting}
/run/beamOn N
\end{lstlisting}

\noindent where \texttt{N} is the number of desired events. In this case, BDSIM
creates 1 run with \texttt{N} events for each time the command is issued.

Geant4 provides several places in the framework where the developer can insert
their own code and gain access to simulation information. BDSIM implements
actions at both the beginning and end of each event and run where
information is collected, histograms prepared and data written to the output.

\subsection{Output}
\label{ssec:output}

A Geant4 simulation produces no output information by default as the total possible
information is unmanageable for storage or analysis. Therefore, it is left to the
developer to provide classes that process the information available during the
simulation from the Geant4 kernel to create the desired reduced output information. BDSIM
uses the Geant4 interfaces and records information from the simulation in two ways.

Firstly, several sensitive detector classes (inheriting \linebreak
\texttt{G4VSensitiveDetector}) are provided and automatically
attached to various volumes. These are registered with Geant4 and are provided
with access to all particle tracks through the volumes they are attached to.
BDSIM includes such a class to record the energy deposition in all volumes
in both 3D Cartesian and curvilinear coordinates.

Secondly, BDSIM records trajectory information independently of volumes at
an event level. As the full set of trajectories for all particles in an event
could reach several gigabytes per event, there are several user options in
BDSIM to downselect which trajectories are desired. Each trajectory consists of
a series of trajectory points that each contain the spatial coordinates of that
point, particle species, total energy and the physics process identification number
associated with the last step. Even though a user may filter to select only a certain
type of particle, the trajectories are stored in a linked manner so that
any individual trajectory point stored can be fully traced back to the primary
particle in the event.

In a conventional tracking simulation the coordinates of each particle are updated
after passing through each element in the accelerator. Therefore, the most
common output is a list of all the particle coordinates after each element.
To be able to compare BDSIM to tracking simulations, a similar functionality
is provided. BDSIM has a \texttt{sample} command that inserts a `sampler' after
either a single specified element or all elements in the beam line. A sampler
is a 1\,nm thick plane that is 5\,m\,$\times$\,5\,m transversely made of vacuum in
a parallel world so as not to interfere with the physical interactions. There is no ability
to record particles on an arbitrary plane or surface of a volume in Geant4, so
a volume must be created and a sensitive detector attached to it. The box is made
as thin as possible to minimise artificially increasing the length of the beam line.
The sampler plane records any particles passing through it.

Output information is stored in the ROOT data format~\cite{ROOT:1997, ROOT:2009}. This is
a well-documented, compressed binary format that is widely used in the high
energy physics community. The ROOT data storage facilities are highly suited to
storing information on an event by event basis and allow direct serialisation of
the developer's C++ classes as well as provision for data schema evolution.
These can be loaded using the compiled software, but the file also
contains a complete template of all classes used such that this is not required
and the data can always be loaded even if the original software is lost.

Although written in compiled C++, the
ROOT framework provides reflection for the classes stored that allows the
compiled code to be easily loaded
and used interactively in the ROOT interpreter as well as in Python with
the exact same functionality allowing users to explore the data interactively
with ease.

Aside from raw information, histograms of energy deposition and primary particle
impact and loss points are recorded with each event. These can be averaged in
analysis to produce the mean energy deposition across all events, but with
the correct statistical uncertainty. This treatment is only possible with
event by event data storage.

Each data file includes information about the BDSIM options used, random
number generator seed states and software identification numbers for BDSIM
and its main dependencies to allow any simulation to be strongly recreated.
Such information allows reproduction of any simulation immediately or even years later.

\subsection{Primary Particle Generation}
\label{ssec:primarygenerator}

Generally an event may start with several initial particles
(`primaries'), however, in the case of an accelerator it is more typical
to simulate a single particle sampled from a beam distribution.

To begin each event, BDSIM draws randomly a set of coordinates ($x$, $x^{\prime}$,
$y$, $y^{\prime}$, $\delta t$, $E$) from a distribution
chosen by the user. BDSIM provides 12 possible distributions. The most basic is
the `reference' distribution where each particle is the same for each event
with a fixed set of coordinates as chosen by the user. These are by default
aligned to the axis of the accelerator with no transverse position or momentum.
This distribution is used to validate the reference or design trajectory.

A 6D Gaussian distribution is provided where the user may specify the standard
deviation $\sigma$ in each dimension as well as the off-diagonal correlation
terms in a 6\,$\times$\,6 matrix. The Twiss parameterisation common to accelerator
beam descriptions is provided, which in turn uses the 6D Gaussian generator.

As one of the purposes of BDSIM is to simulate lost particles interacting
with the accelerator, which usually occurs with a low frequency,
several distributions are provided that make the simulation more efficient.
One such distribution is the `halo' distribution that provides a
particle distribution according to the nominal Gaussian beam, but at a large
number of $\sigma$.

In all cases, the generator uses a single instance of the pseudo-random
number generator from the CLHEP library. The user may specify a starting
seed value or one is automatically generated from the computer clock.

\begin{figure*}
  \normalsize
  \centering
  \includegraphics[width=0.9\textwidth]{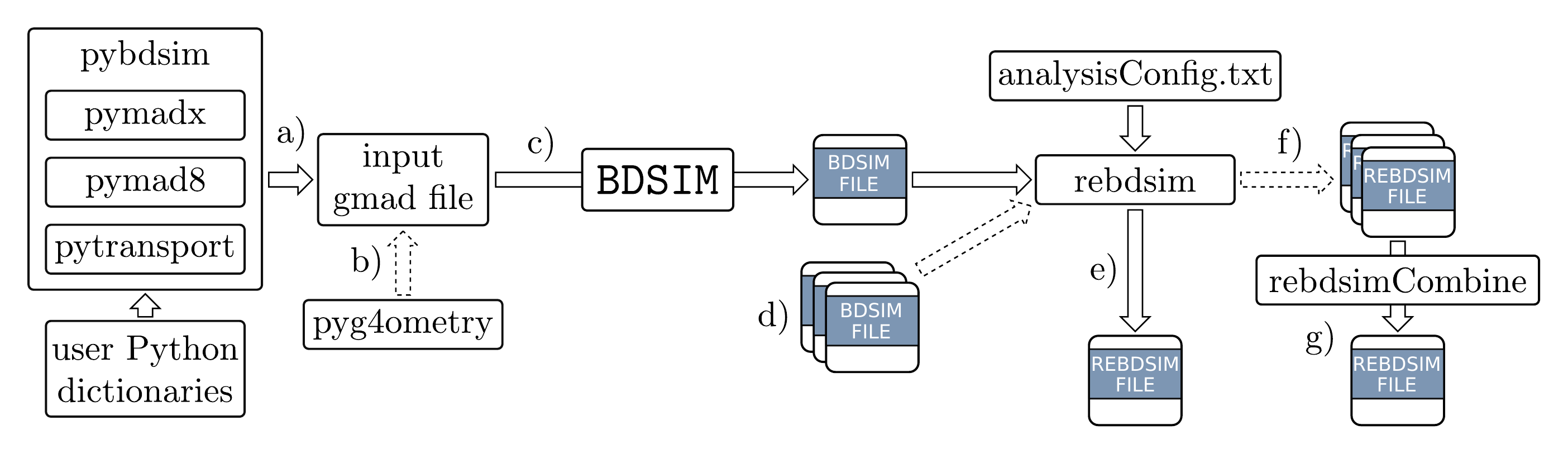}
  \caption{\label{fig:workflow}Diagram showing the workflow using the extended tools included with
    BDSIM. The dashed arrows represent optional steps. From left to right: a) \texttt{pybdsim} is used
    in combination with other Python packages to prepare an input model including optional input user-provided
    information in Python dictionaries, b) \texttt{pyg4ometry} can optionally be used to help prepare input
    geometry, c) BDSIM is used to generate data, d) BDSIM is used many times with multiple seeds on a computer
    cluster to generate many output files for a study, e) \texttt{rebdsim} is used to analysed one or multiple
    output files at once using an analysis configuration text file, f) optionally, \texttt{rebdsim} is used
    one-to-one on output files directly producing one \texttt{rebdsim} file per BDSIM file, in which case
    g) \texttt{rebdsimCombine} is used to combine these into one result.}
\end{figure*}

\subsection{Simulation Control}
\label{ssec:control}

In each event, particles are tracked until they reach zero kinetic energy or
they leave the world (outermost) volume. With high energy primary particles,
a very large number of secondaries can be produced and as the particles decrease
in energy the number of secondaries can grow, which is known as infrared divergence.
Such detail at low energy may not be necessary and may dominate computation time.
To avoid this, it is necessary to have some control over the physics processes
other than including only those relevant in the physics list.

Geant4 provides the ability to provide tracking cuts in a volume. Here, a minimum
energy and maximum time can be specified. If a particle has an energy under or a
time greater than these limits, the particle will be removed from tracking. BDSIM
provides a method to set limits that will be attached to all volumes and also
records the final energy of the particles to conserve energy per event.

In addition to the tracking cuts, Geant4's primary mechanism is a range cut. A
range cut is a distance assigned to a particle species that a
secondary of that species would be required to travel. If the secondary would not
travel at least that range, the energy deposition is recorded but the secondary
particle is not produced. The range is internally converted in Geant4
to a material and particle specific energy cut. This strategy provides the
greatest accuracy with the least computation~\cite{GEANT4:2003}. BDSIM provides
an interface to set the range cut for protons, electrons and positrons, photons
as well as a default.

Such limits are often necessary as a simulation may not include physics processes
that would lead to the natural termination of a particle. For example,
synchrotron radiation can produce a very large number of photons making
it computationally expensive to track all the secondary particles. The rate
scales $\sim\,E^4$, so it can dominate when simulating high energy events and
therefore is often omitted. This omission however, may lead to low energy particles
spiralling in a magnetic field, such as that of a dipole, indefinitely. BDSIM's
numerical integrators have special treatment of spiralling particles, but the
user limits provide a convenient method to avoid such scenarios that may lead
to long running events with no gain in information.

\begin{sloppypar}
One further consideration is a circular accelerator. With a circular accelerator
and no synchrotron radiation, the particle energy will not decay as the magnetic
field does no work. BDSIM therefore provides a \texttt{circular} option to limit the
number of turns any primary particle can complete. A
5\,m\,$\times$\,5\,m\,$\times$\,0.1\,$\mathrm{\mu}$m cubical volume is inserted
between the beginning and end
of the lattice that is orientated to make a large transverse plane to the beam.
A dynamic set of user limits is attached as well as a special
Geant4 sensitive detector class. The sensitive detector class records the
number of turns completed by the primary particle and when this has reached the
specified number of turns, the user limits are dynamically changed to reject
any particle over 0\,eV. This makes the box an infinite absorber, stopping
any particle that hits it.
\end{sloppypar}

\subsection{Variance Reduction}
\label{ssec:variancereduction}

With the large number of physics processes included in each physics list
as well as the large number of possible outcomes from each interaction,
the outcome or process of interest may occur at a low frequency per event.
An efficient simulation would simulate the outcome one wishes to characterise
for each event, i.e. it is impractical to simulate tens of millions of events
to observe only a few occurrences of the desired outcome. When analysing
a data set, this effect leads to a large variance of results when in
certain areas of parameter space. Biasing is a form of variance reduction
where a process or outcome is made artificially more frequent, but
recorded with a corresponding statistical weight.
Multiplying
the simulated frequency by the weight gives the correct physical result but,
due to the more frequent occurrence, with a reduced variance. Traditionally,
the developer had to write their own C++ wrapper for a given process they
wish to study, but recent developments of Geant4 have introduced a more
generic biasing interface~\cite{GEANT4:2016}. Geant4's generic biasing
interface provides both physics based (process cross-section) and non
physics based (splitting and killing) such as geometrical importance
biasing or importance sampling. An interface to process biasing
is provided in BDSIM that allows the user to scale the process
cross-section by a given factor for a given particle. The biasing
can be applied globally, or to specific \emph{vacuum} volumes
(inside the beam pipe) or the general material (anything outside
the beam pipe) of specific beam line elements. For example, this can
allow simulation of elastic or inelastic interaction of particles with
residual gas molecules in the `vacuum' of the beam pipe.

\subsection{Visualisation}
\label{ssec:visualisation}

At the initial stages of a study, it is crucial to visualise a model
to validate its preparation as well as the expected outcome of a typical
event. Visualisation is accomplished through an interface to the Geant4
visualisation system that provides a variety of different visualisation
programs depending on the software environment.

The default and recommended visualiser is the Geant4 Qt visualiser. This
provides a rich interface with an interactive 3D visualisation where the
model can be viewed with solid surfaces or as a wire-frame, both with and without
perspective. A built in command prompt allows extensive control of the
simulation and the visualisation, such as geometry overlap checking and
particle track colour schemes.

To visualise an event, all trajectories must be stored. Additionally, the
model and all trajectories must be converted to polygon meshes and rendered
on screen. Whilst this is handled by the Geant4 visualisation system, this
overhead leads to the interactive visualisation system running events
approximately an order of magnitude slower than they would without visualisation.
The visualisation is crucial in the first stages of a simulation for validation
but beyond that it is more desirable to generate as many events as possible
in a given time. BDSIM by default will run interactively, but if executed with
the \texttt{--batch} option, it will not use the visualiser and complete the
simulation with only textual diagnostic and informative output to the terminal.
Batch mode is much faster and suitable for simulations that may be run on a
computing farm where no graphics systems are available.

\subsection{Analysis}
\label{ssec:analysis}

Once BDSIM has produced output, this may be analysed using an included
analysis suite. This covers the most common basic analysis but also
includes an interface for the user to include more complicated analyses.

The main analysis tool is \texttt{rebdsim} (`ROOT event BDSIM'). This uses
a simple text file as input that defines histograms to make from the
output structures contained in output files. These can be one to three dimensional
histograms made on an event by event basis or as a simple integration across
all events. Any histograms stored in the raw output that were produced in
BDSIM, such as energy deposition and primary impact location, are combined
to produce a mean histogram across all events. The histogram definition in
\texttt{rebdsim} specifies which variables in which `Tree' in the output to make the
histogram from as well as binning and a `selection'. The binning can also be
chosen to be logarithmic to cover a large range in values. The selection is an
optional weighting that can be a numerical factor, another variable in the data,
a Boolean expression based on data variables or a combination of all. This
is an interface to that of \texttt{TTree:Draw} in ROOT.

The event level structure in the output is paramount to the analysis for two reasons

\begin{enumerate}
\item To calculate the correct statistical uncertainties.
\item To filter independent events on any variable.
\end{enumerate}

To correctly calculate the variance and therefore the statistical uncertainties
of any mean histograms, the data must be structured in a per event manner.
Conventional tracking programs only simulate one particle per `event' so
there is no need to structure data in this way and the uncertainties are trivially
calculated. However, with the complex
tree of particles and physics processes that can happen per event in a
radiation simulation, it is absolutely crucial to structure the data in this way. The
default histograms made by \texttt{rebdsim} are made event-by-event and so have the
correct statistical uncertainties. The ability to use a selection and filter events
based on any variable permits any user analysis to be performed without prior
knowledge of the beam line or physical origin of particular information and is
highly powerful.

Other attempts at similar simulations typically keep track of very specific
information for that application limiting the usefullness of the code. None
of the comparable codes already described preserve a per event structure in
data and deal only with simple integral histograms and therefore lack
the correct treatment of statistical uncertainties.

\begin{sloppypar}
Using the `chain' feature of ROOT many files can be analysed together
behaving as one. Furthermore, a tool \linebreak \texttt{rebdsimCombine} is provided to
combine the output from several instances of \texttt{rebdsim}. For large data sets
it would be prohibitive to analyse the whole data set serially, so it is
preferable to analyse small chunks in parallel and then combine the resultant
histograms. This strategy results in the exact same numerical answer but
in a fraction of the time.
\end{sloppypar}

The ROOT output format and included analysis tools provide great flexibility
in the storage and analysis of simulation data. They also ensure the process
is scalable to the very largest data sets dealt with today on the multi-terabyte
level.

\subsection{Workflow}
\label{ssec:workflow}

Apart from the main program, BDSIM, a suite of associated tools is included to
facilitate a smooth workflow. These aid in input automatic input preparation
and analysis and combination of results as shown in Figure~\ref{fig:workflow}.

To aid input preparation, several Python packages are included that aid with different
input formats as described in Table~\ref{tab:pythonutils}. There are individual
packages for common output formats from accelerator design software such as
\MADX{}~\cite{MADX-website}, MAD8~\cite{MAD8}, and TRANSPORT~\cite{TRANSPORT, TRANSPORT-modern}
(\texttt{pymadx}, \texttt{pymad8}, and \linebreak \texttt{pytransport} respectively).
These can be used individually to load, inspect and plot information, but
are also tied together by the more general package \texttt{pybdsim} that allows conversion
from these formats to a BDSIM model. The conversion includes the ability to include
extra information (via user-provided Python dictionaries) in the converted model
that may come from other sources, such as aperture or collimator information.

A further tool, \texttt{pyg4ometry}~\cite{pyg4ometry-website, pyg4ometry-manual-website}, can be
used to prepare GDML geometry files for use in BDSIM or any Geant4 simulation. It can
also be used to aid in the conversion from other geometry formats. Python scripts
can be used to make simple components that can be easily included in a beam
line in BDSIM.

\begin{table}
  \centering
  \caption{\label{tab:pythonutils} Included Python packages with BDSIM.}
  \begin{tabular}{l p{6cm}}
    \hline \hline
    Package              & Description                               \\ \hline
    \texttt{pymadx}      & Load MAD-X output and create MAD-X models \\ 
    \texttt{pymad8}      & Load MAD8 output and create MAD8 models   \\ 
    \texttt{pytransport} & Parse TRANSPORT input                     \\ 
    \texttt{pybdsim}     & Convert model using above packages, load and plot raw
    and analysed BDSIM data                                         \\ \hline \hline
  \end{tabular}
\end{table}

After running BDSIM, the output is a ROOT file with per-event data. This requires
analysis as defined by the user to reach a result. Accompanying BDSIM are several
other programs that aid with analysis and post-processing of data and are described
in Table~\ref{tab:otherprograms}.

\begin{table}
  \centering
  \caption{\label{tab:otherprograms} Included programs other than BDSIM.}
  \begin{tabular}{l p{4.7cm}}
    \hline \hline
    Program                     & Description                                              \\ \hline 
    \texttt{rebdsim}            & Analyse raw BDSIM data files producing a histogram file  \\ 
    \texttt{rebdsimCombine}     & Combine multiple histogram files from \texttt{rebdsim}   \\ 
    \texttt{rebdsimHistoMerge}  & Combine only pre-made energy deposition histograms from
                                  a set of BDSIM data files                                \\ 
    \texttt{rebdsimOrbit}       & Extract the first hit in each sampler to easily access
                                  the primary particle at each sampler                   \\ 
    \texttt{bdsinterpolator}    & Load BDSIM format field maps, interpolate and export   \\ 
    \texttt{comparator}         & Regression testing tool for comparison of data         \\ 
    \texttt{ptc2bdsim}          & Convert MAD-X PTC ASCII output to a BDSIM data file for
                                  easier comparison with BDSIM data                      \\ \hline \hline
  \end{tabular}
\end{table}

\section{Accelerator Tracking}
\label{sec:accelerator-tracking}

The Geant4 model constructed by BDSIM provides all the required fields
for an accelerator, however, the commonly used numerical integration
algorithms provided with Geant4 such as a 4$^{th}$ order Runge-Kutta integrator will
not provide sufficient accuracy for tracking particles through an accelerator.
Whilst these numerical integrators are suitable for arbitrary spatially and time
varying magnetic and electric fields, the small errors from
the numerical integration can accumulate with many successive uses leading to
inaccurate results, hence
these are rarely used for accelerator tracking. For the specific static
magnetic fields of an accelerator, such as a pure dipole or quadrupole
field, there are exact solutions that provide more accurate tracking.
However, these typically use a coordinate system that follows the
accelerator. Algorithms for these tracking routines as well as
coordinate system transforms are included with BDSIM and used by default.

\begin{figure}[htb!]
  \centering
  \includegraphics[width=7cm]{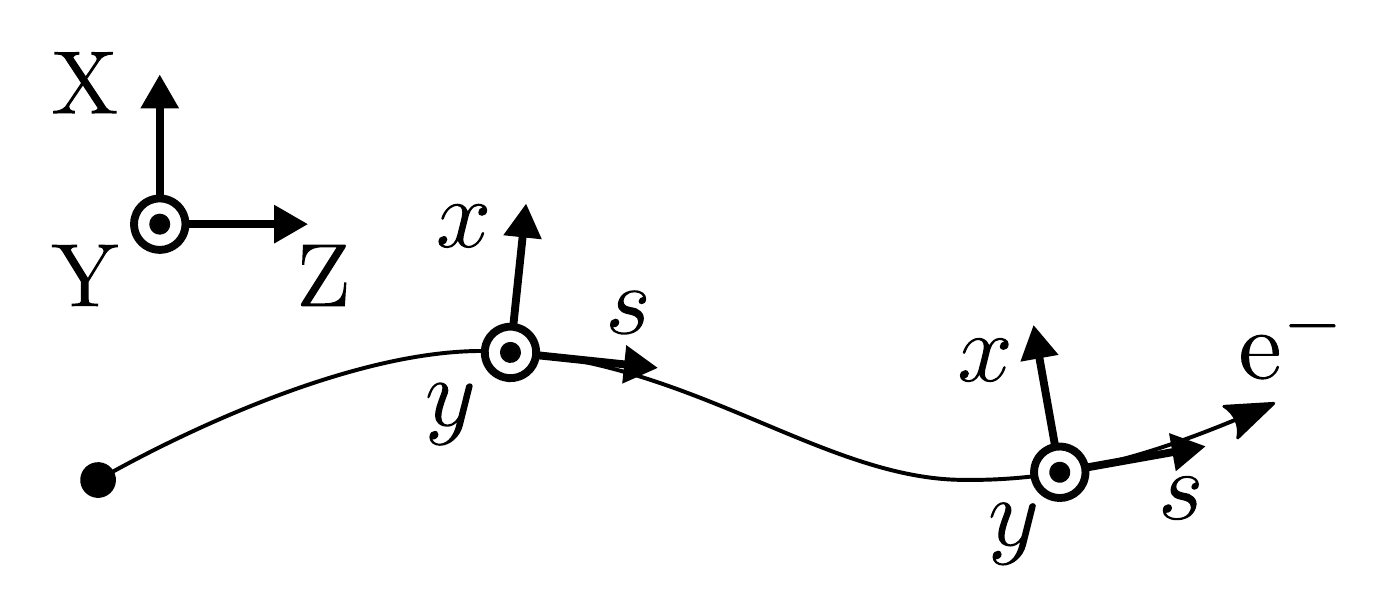}
  \caption{\label{fig:frenet-serret}The curvilinear Frenet-Serret coordinate system
    ($x$, $y$, $s$) shown following the trajectory of the reference particle.
    The global Cartesian system is shown (X, Y, Z).}
\end{figure}

The majority of accelerator tracking algorithms use a coordinate system
that follows the path of the reference particle (no transverse momentum
and exactly the design energy) as opposed to 3D Cartesian
coordinates~\cite{Wiedemann:2007}. The Frenet-Serret is such a curvilinear
coordinate system as shown in Figure~\ref{fig:frenet-serret}.

With the Frenet-Serret coordinate system each element's numerical accuracy is
independent from its location and there are no repeated transformations
into and out of the system the algorithms are specified in.

Accelerator tracking algorithms can broadly be classed in two categories:
thick, and thin. In the thick regime, elements are as long as they are in
reality. In the thin regime, a series of instantaneous `kicks', i.e momentum changes,
interspersed with drifts are applied to the trajectory. The kicks affect only the
transverse momentum and not the position
of the particle. As expected, a single kick is not a
physically accurate representation of the passage of a particle through a
magnetic field, however, the weighted combination of many small kicks in
combination with drifts does~\cite{Schachinger}. The thin regime is often used as it
is more computationally performant for a solution that conserves area or volume
in phase space. A solution that conserves phase space volume (i.e. \emph{symplectic})
is required as the tracking algorithms may be applied many
times and any small errors will accumulate until the result is no longer
physically correct. General symplectic thick solutions exist, but are often
more computationally expensive. An accelerator `lattice' is typically first
described by a thick lattice that matches the physical accelerator and
then potentially converted to a thin lattice.

Geant4 provides a series of numerical integrators that are supplied with
spatial coordinates and the field values at that location and are required
to predict the particle motion. These must be able to handle a variety of
situations that are not encountered in an accelerator tracking program:

\begin{itemize}
\item An arbitrary step length.
\item Particles with different masses.
\item Particles with different charge or that are neutral.
\item Backwards travelling particles.
\item Particles travelling parallel to a magnetic field.
\end{itemize}

To introduce accurate and robust accelerator tracking to a Geant4 model, any new
integrator must be able to handle these situations. Furthermore, the
3D nature of the model prevents the representation of the accelerator
as thin lenses. As Geant4 assumes numerical integration, any new integrator
must also estimate a numerical uncertainty associated with the calculation.

BDSIM includes a set of integrators that encapsulate thick tracking algorithms
that use the Frenet-Serret coordinate system with special provision for
the aforementioned tracking scenarios that may occur in a Geant4 model.
These integrators ignore the supplied field from Geant and make use of the
normalised strength they are constructed with (e.g. $k_1$ for a quadrupole).
For dipole and quadrupole fields, an exact analytical solution is possible
for the particle motion and numerical integration is not required. For higher
order magnets such as sextupoles or octupoles, 2$^{nd}$ order Euler integrators
are provided.

The required coordinate transforms between 3D Cartesian and the Frenet-Serret
system make use of the parallel geometry constructed by BDSIM. For scenarios
where these algorithms cannot be used, a 4$^{th}$ order Runge-Kutta integrator
is used. It is foreseen that no one set of algorithms may be applicable
to all situations, or they may be extended in future, so sets of integrators
are used. The user may select from different predefined sets depending on
the application. The default set is designed to be both physically accurate
and match optical design tools such as \MADX{}.

A full mathematical description of all BDSIM integrators is given in the
provided documentation~\cite{BDSIM-manual}.

For the most accurate model, the user may choose to use an electromagnetic field map from
simulation or measurement. Such field maps can be loaded by BDSIM and overlaid
on to both BDSIM and user-provided geometry replacing default fields. The supplied
field maps can be
easily converted from other formats using the included \texttt{pybdsim} Python utility
and can be compressed using gzip. The gzstream library~\cite{gzstream-website}
is used to dynamically decompress the text files as they are loaded.
BDSIM includes a variety of
interpolators to provide the continuous field values in 3D coordinates
required by the simulation from the field specified at discrete points in the
field map. The user may choose any of the Geant4 numerical integrators
ranging from low order Euler to 8$^{th}$ order Runge-Kutta ones.

This approach allows a user to start with a generic model that will match
the design optical description and then further customise with specific
field maps.

\section{Example}
\label{sec:example}

\begin{figure}
  \centering
  \includegraphics[width=8.5cm]{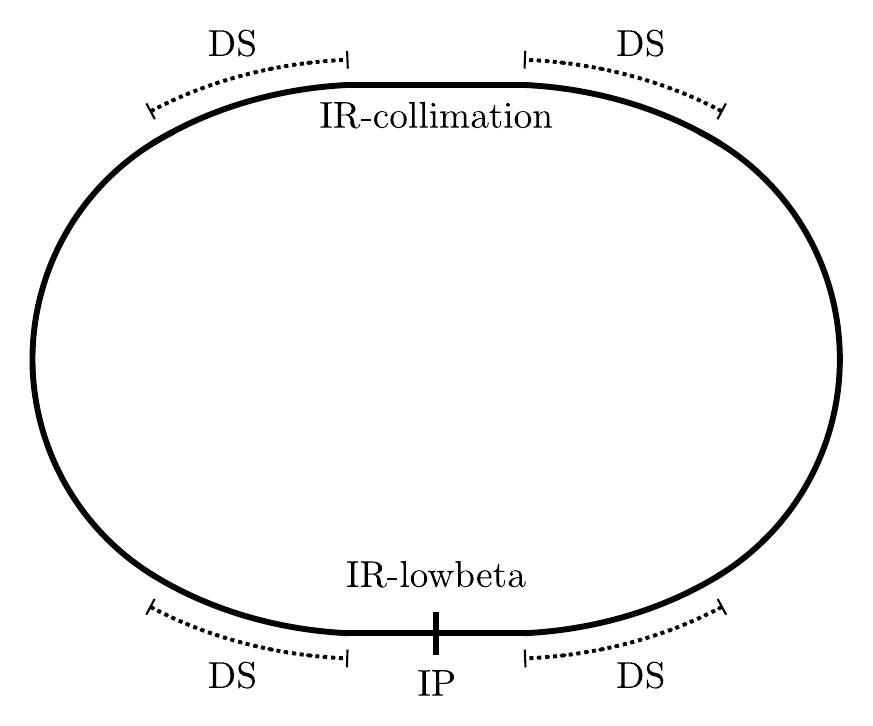}
  \caption{\label{fig:examplelayout} Layout of the example model with race
    track design including two arc sections, and two insertion regions: `IR-collimation'
    and `IR-lowbeta'. Between each arc and an insertion region there is a
    dispersion suppressor `DS'. The interaction point `IP' is where the beam is
    focussed to a minimum size.}
\end{figure}

To demonstrate the capabilities and unique features of BDSIM, an example
model is provided and illustrated here. The model is of a fictional
accelerator that demonstrates many features found in a variety of
real accelerators. The model consists of a 4\,km racetrack ring
with two straight sections that include a low-$\beta$ collision point
and a high beta collimation section. Each insertion has a 3-cell
half-dipole-strength dispersion suppressor on each side. The most
relevant parameters are shown in Table~\ref{tab:modelmodelparams}. The model
and all materials required to prepare it are included with BDSIM
as a worked example called `model-model'. A schematic layout
is shown in Figure~\ref{fig:examplelayout}.

\begin{table}
  \centering
  \caption{\label{tab:modelmodelparams} Defining parameters of example model.}
  \begin{tabular}{l l}
    \hline \hline
    Parameter             & Value       \\ \hline
    Particle              & proton      \\ 
    Energy                & 100\,GeV    \\ 
    Length                & 4\,km       \\ 
    Dipole bending radius & 221.5\,m    \\ 
    $\epsilon_{N\,x,y}$    & 1\,mm\,mrad \\ 
    Collimators           & 10          \\ 
    Dipoles               & 256         \\ 
    Quadrupoles           & 252         \\ 
    $\beta^{\ast}_{x,y}\*$ & 2.517\,m     \\ \hline \hline
  \end{tabular}
\end{table}

\begin{figure*}
  \normalsize
  \centering
  \includegraphics[width=0.95\textwidth]{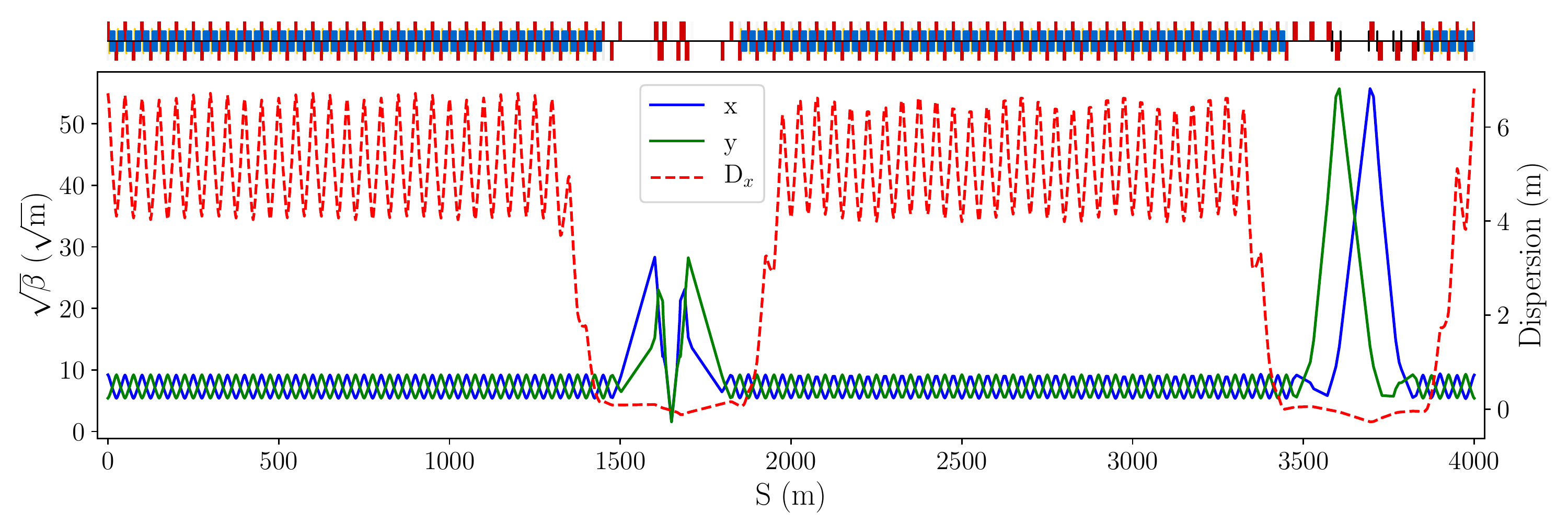}
  \caption{\label{fig:modeloptics}\MADX{} linear lattice functions of the example model. The two
    insertion regions can be seen with reduced horizontal dispersion (\emph{red dashed}).
    The first insertion creates a small symmetric beam suitable for collision
    with another beam or gas target. The second insertion produces a large beam size suitable
    for collimation. The schematic at the top of the plot depicts each magnet
    type in the accelerator (blue: dipoles; red: quadrupoles, yellow: sextupoles,
    black: collimator).}
\end{figure*}

The model was created in \MADX{}~\cite{MADX-website} and optical functions as calculated
by \MADX{} are shown in Figure~\ref{fig:modeloptics}. The low-$\beta$ straight section is
designed to
create a small symmetric focus of the beam suitable for a collision point with
potentially another beam or gas target. The other insertion is designed to expand
the beam suitable for collimation or scraping of the beam.

\begin{figure}
  \normalsize
  \centering
  \includegraphics[width=8.5cm]{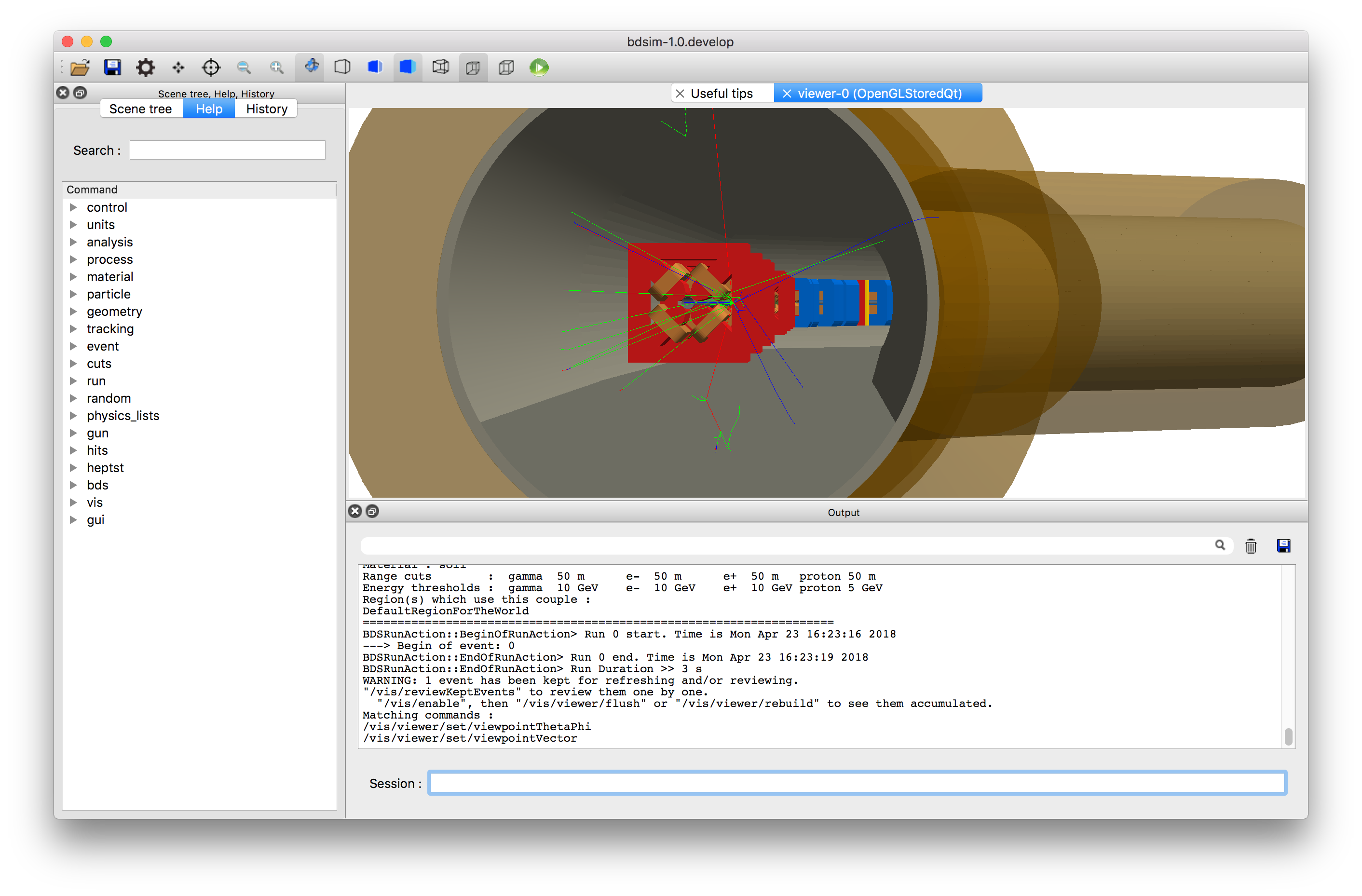}
  \caption{\label{fig:modelvisualisation}Visualisation of the example model in BDSIM
    using the Geant4 Qt visualiser. Only part of the machine is visualised showing a
    single proton impact and secondaries (colour coded by charge). The optional tunnel
    is automatically built by BDSIM to follow the beam line.}
\end{figure}

\begin{figure}
  \normalsize
  \centering
  \includegraphics[width=8.5cm]{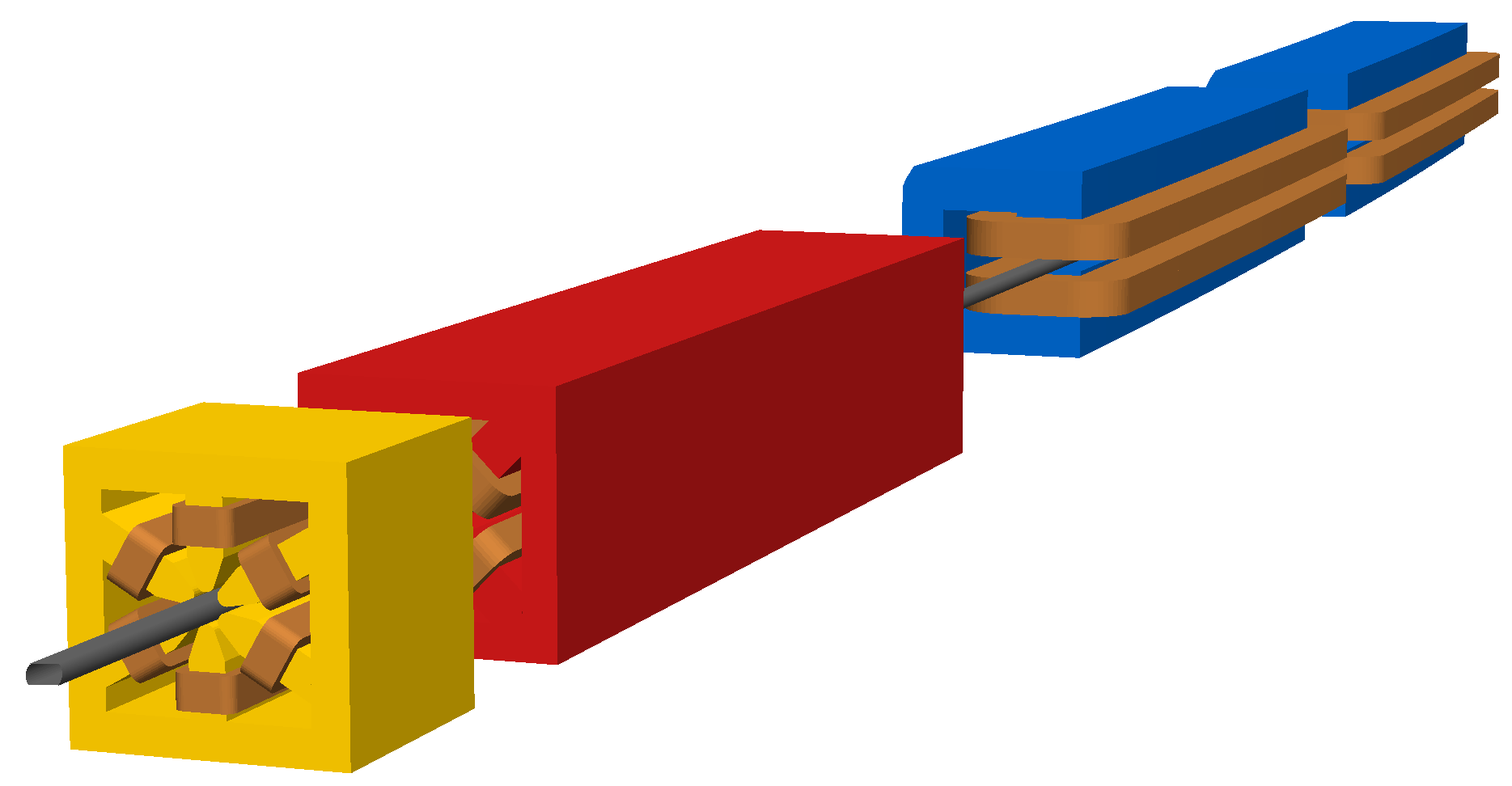}
  \caption{\label{fig:modelvisualisation2}Visualisation of four magnets from the
    unit cell in the arcs of the example model in BDSIM. From left to right, a
    sextupole, quadrupole and two sector bends (dipoles) can be seen using the
    generic geometry provided by BDSIM.}
\end{figure}

To prepare a BDSIM model an output version is first
made from \MADX{} using the TWISS command that produces a sequential one-to-one
representation of the machine in an ASCII file.
This is trivially converted to BDSIM input format GMAD using a provided Python
converter \texttt{pybdsim}. Any additional information not specified in this file
such as aperture or collimation settings can be included in this conversion with
user-supplied Python dictionaries. There is no standard format of auxiliary information
about an accelerator so loading this information is left to the user, however
Python is a widely used language for which many data loading and processing libraries
exist. The automatically converted model includes a Gaussian beam distribution as
parameterised by the Twiss parameters from the first element in the sequence and beam information
from the header. This fully
functional BDSIM model created in minutes acts as a starting point that can then be customised.
Figure~\ref{fig:modelvisualisation} shows the 3D visualisation of part of the model in BDSIM and
Figure~\ref{fig:modelvisualisation2} shows the magnets from part of a unit cell in the arc.

This \MADX{} model includes only basic information about the collimators such as their
length and perhaps an aperture description, however excludes material definition.
The extra required information is provided in an included text file that specifies
collimator openings as a specific number of $\sigma$ in the beam distribution at that
point as well as material definition. This is used as an example of including
extra information into the model conversion and preparation process that could
come from an external source.
Table~\ref{tab:collimatorsigmas} illustrates the general settings used in the 3-stage
collimation system that has `primary', `secondary' and `tertiary' collimators that
are placed at successively greater distances from the beam. Each collimator consists
of a rectangular aperture with a narrow opening in one dimension and a more open
one in the other that is designed not to impede the beam.

\begin{table}
  \centering
  \caption{\label{tab:collimatorsigmas} Collimator openings in beam $\sigma$. The
  `open' setting is used for the aperture in the non-collimating plane.}
  \begin{tabular}{l c}
    \hline \hline
    Collimator Type    & $\sigma$ opening    \\ \hline
    primary            & 6                   \\ 
    secondary          & 7                   \\ 
    tertiary           & 8                   \\ 
    open               & 40                  \\ \hline \hline
  \end{tabular}
\end{table}

These conceptual settings are used in combination with the \MADX{} optical functions
to calculate absolute collimator openings in millimetres using an included Python
script. These are shown in Table~\ref{tab:collimatorgaps}.

\begin{table}
  \centering
  \caption{\label{tab:collimatorgaps}Collimator half openings as calculated from
    a given beam $\sigma$ and the optical functions at each collimator location.
    Primary collimators are composed of carbon, secondary collimators copper and
    tertiary absorbers tungsten that are typical of high energy proton collimation
    systems with increasing atomic Z.}
  \begin{tabular}{l c c c}
    \hline \hline
    Name        & Material  & $\Delta_{x}$ (cm) & $\Delta_{y}$ (cm) \\ \hline
    COLPRIMY    & C         & 3.174            & 2.660 \\ 
    COLSECY     & Cu        & 5.891            & 3.681 \\ 
    COLPRIMX    & C         & 3.133            & 6.066 \\ 
    COLSECX     & Cu        & 3.157            & 3.250 \\ 
    COLTERT     & W         & 1.434            & 4.437 \\ 
    COLTERT2    & W         & 0.801            & 0.613 \\ 
    COLTERT3    & W         & 0.523            & 0.584 \\ \hline \hline
  \end{tabular}
\end{table}

The model is prepared using the included \texttt{pybdsim} converter. This allows the
extra information for collimators to be included as a Python dictionary that is
prepared from the text file. The converter also allows the inclusion of aperture
information loaded from a \MADX{} TFS file using \texttt{pybdsim}. The classes for aperture
loading in \texttt{pybdsim} also permit aperture filtering and substitution. The conversion
is shown below where `cols' is a Python dictionary of collimator information by name.

\begin{lstlisting}[language=Python]
import pybdsim
import pymadx
ap = pymadx.Data.Aperture(`ring_aperture.tfs')
ap = ap.RemoveBelowValue(0.005)
cols = pybdsim.Data.Load(`collimatorSettings.dat')
pybdsim.Convert.MadxTfs2Gmad(`ring.tfs', `bdsim-model', aperturedict=ap, collimatordict=cols)
\end{lstlisting}

The user may use the converted model immediately, edit it to their needs or include
the automatically produced GMAD files in their own models. Including the files in
a user-written model allows the model to be safely regenerated at any point without losing any
user-defined input.

After preparation of a model, the first step is to validate the optical functions
of the BDSIM model to ensure correct preparation. To generate optical functions, the
automatically provided Gaussian beam distribution according to the Twiss parameters
at the start of the machine is used. The beam distribution is sampled after each
element in the BDSIM model by including the sampler command in the input gmad:

\begin{lstlisting}
sample, all;
\end{lstlisting}

To estimate the optical functions, approximately 1\textendash{}10 thousand particles
should be simulated. The model is run with the command:

\begin{lstlisting}
bdsim --file=bdsim-model.gmad --outfile=optics1 --batch --ngenerate=1000 --circular
\end{lstlisting}

The output ROOT format file `optics1.root' is analysed by an included tool
\texttt{rebdsimOptics} that calculates the optical functions as well as the associated
statistical uncertainty from the finite population at each sampler is also
calculated.

\begin{lstlisting}
rebdsimOptics optics1.root opticalfunctions.root
\end{lstlisting}

\begin{figure*}
  \normalsize
  \centering
  \includegraphics[width=0.95\textwidth]{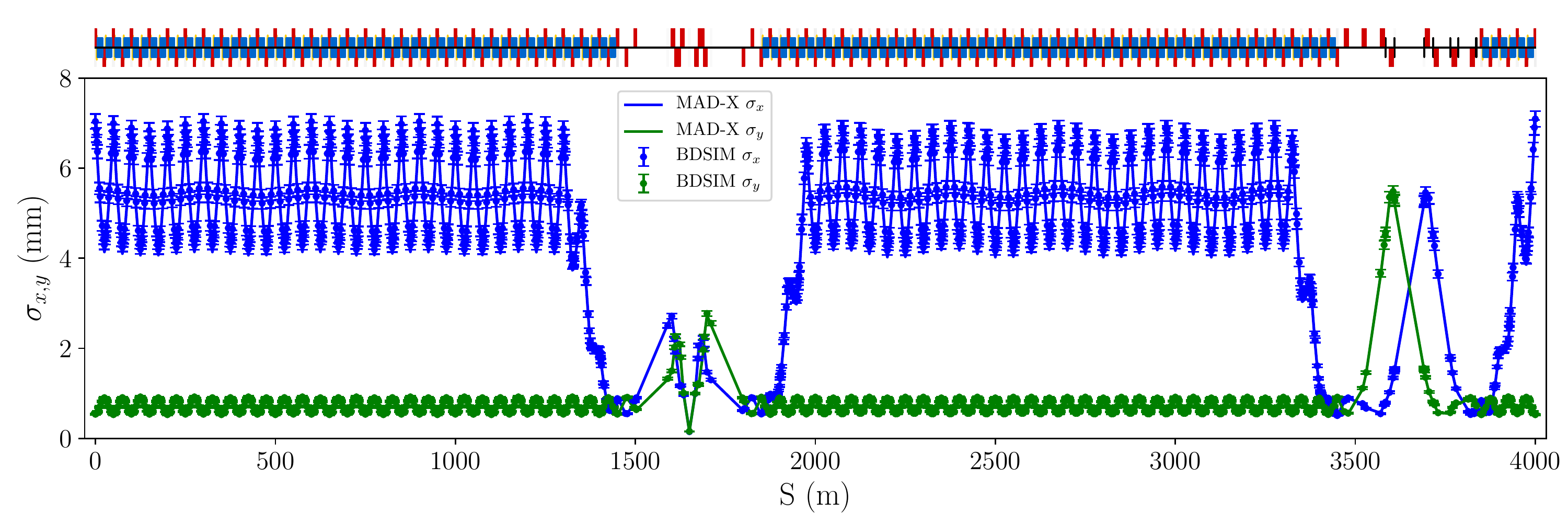}  
  \caption{\label{fig:opticscomparison}Example machine optical comparison of beam size calculated from
    \MADX{} optical functions, emittance and energy spread with that simulated by BDSIM
    showing agreement. The beam size in the arc sections is dominated by the dispersion
    contribution. The schematic at the top of the plot depicts each magnet type in the accelerator
    and follows the sequence around the ring.}
\end{figure*}

These are
calculated by accumulating 1\textendash{}4$^{th}$ order power sums and calculating various
moments of the distribution. The statistical uncertainty reduces with the number
of particles simulated and it is recommended to simulate at least 1000 for a
meaningful comparison. \texttt{rebdsimOptics} produces another ROOT format file
with the optical function data. The optical functions as calculated from BDSIM
can be compared using \texttt{pybdsim}. In this case, we compare against the original
\MADX{} optical functions the model was prepared from using the following command:

\begin{lstlisting}
pybdsim.Compare.MadxVsBDSIM(`ring.tfs', `opticalfunctions.root')
\end{lstlisting}

This produces a series of plots that compare $\bar{x}$, $\bar{y}$, $\sigma_{x, y}$,
$\sigma_{x', y'}$, $D_{x,x'}$, $D_{y,y'}$, $\alpha_{x, y}$ and $\beta_{x, y}$, as well
as the transmission. An example is shown in Figure~\ref{fig:opticscomparison}.

When simulating the optical functions, ideally no particles are lost as this would
affect the shape of the beam distribution and the validity of comparing $\sigma$ of
the distribution. However, once validated and a physics study is desired, the lack of
beam loss makes the simulation very inefficient. For example, to witness an event at
6\,$\sigma$, approximately 500 million events would be required on average to be simulated.
Therefore, it is logical to select a distribution of primary particles that will
collide with the accelerator or exhibit the desired interaction. BDSIM includes a
variety of bunch distribution generators for this purpose.

\begin{figure*}
  \centering
  \includegraphics[width=0.95\textwidth]{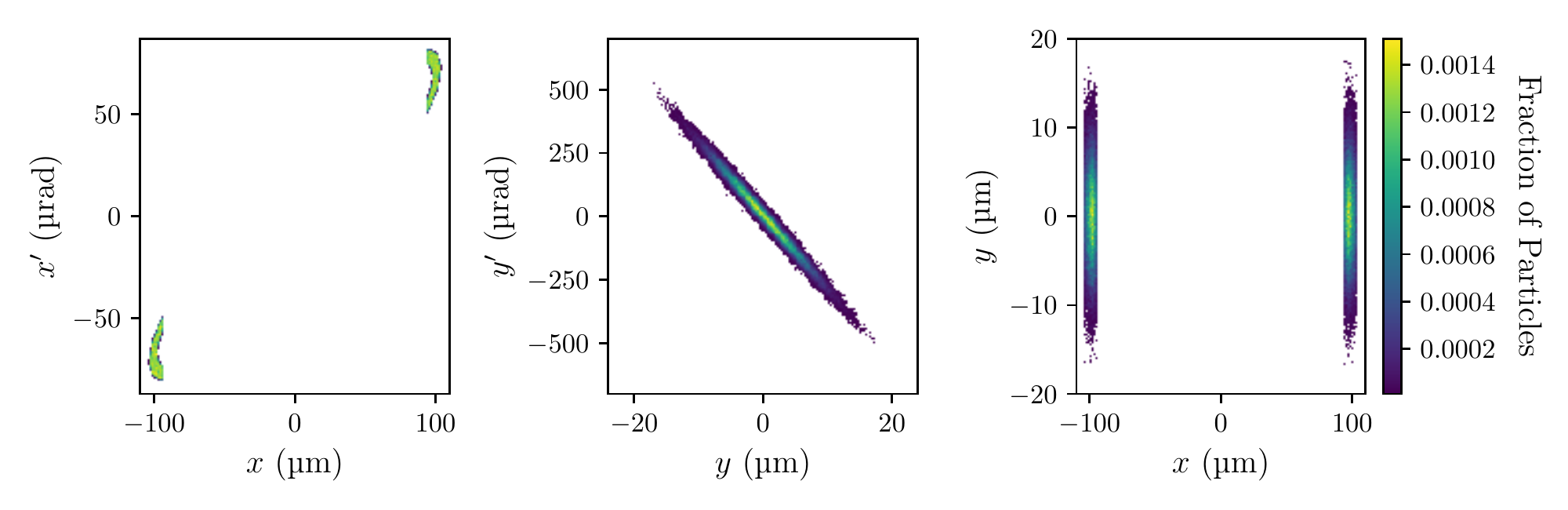}
  \caption{\label{fig:phasespace}Example phase space distribution for a beam loss simulation
    using the `halo' distribution based on the Twiss parameters at the entrance to the
    primary horizontal collimator.
    The horizontal phase space (\textit{left}) is designed to intercept a horizontally
    orientated 2 jaw collimator, whereas the vertical phase space (\textit{centre}) is
    the unaltered distribution according to the Twiss parameters. The combination of the
    two spatially is shown (\textit{right}). The distribution is exaggerated here with
    a greater spread in $\sigma$ than is required to make the shape clear. The horizontal
    phase space is clipped such that only particles with $x > 5.7\sigma$ are produced.}
\end{figure*}

For this example, the intent is to evaluate
the performance of the collimation system. Any beam that exists grossly outside the
collimator aperture will be immediately lost within one revolution in a circular machine.
Similarly, any particle largely inside the collimator aperture will not intercept the
collimator. We therefore simulate a thin annulus in phase space that aligns with the edge
of the collimator closest to the beam. Such a distribution is provided by the `halo'
distribution in BDSIM. This generates particles uniformly in a given phase space volume and
accepts or rejects the particle depending on its single particle emittance.
Furthermore, spatial limits are provided that allow the phase space to be reduced.
An example input distribution is shown in Figure~\ref{fig:phasespace}.

\begin{figure*}
  \normalsize
  \centering
  \includegraphics[width=0.98\textwidth]{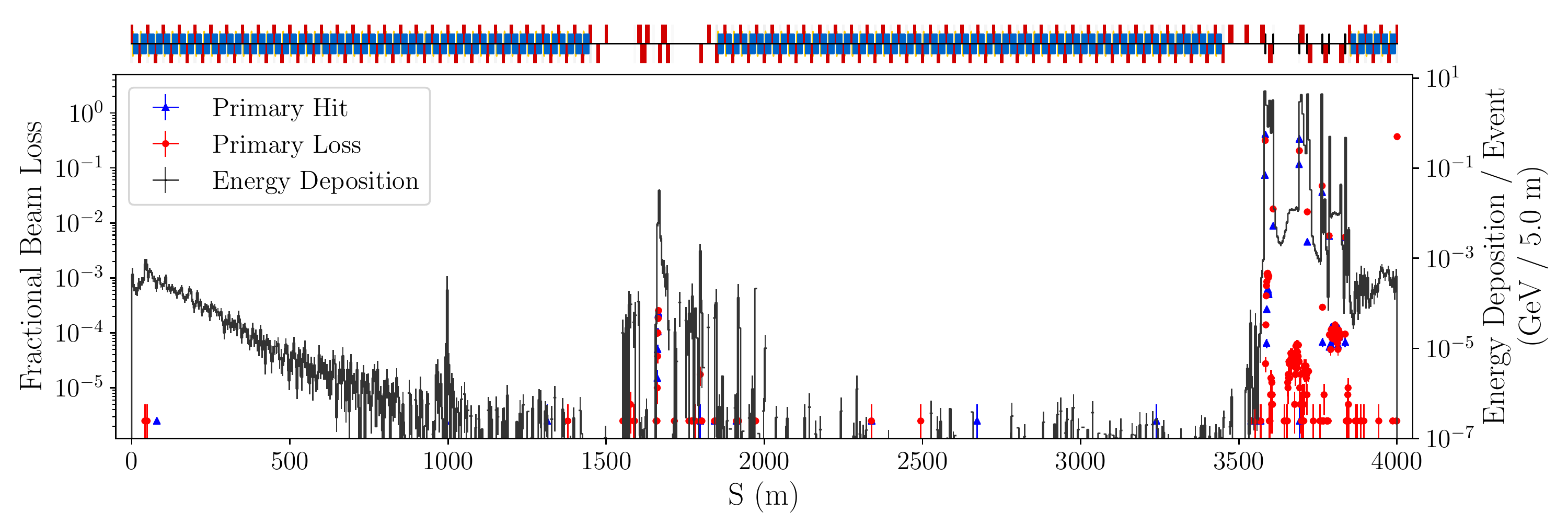}
  \caption{\label{fig:modellosses}Combined beam loss from a horizontal and vertical
    halo simulation with $4\times10^{5}$ primary protons at 100\,GeV as a function of
    distance, S along the axis of the machine. Blue triangles show
    the point of first contact of primary particles and red circles, their stopping
    location. The integrated energy deposition is also shown in black recorded
    by the accelerator geometry. The diagram on top shows the various magnets in order. The
    beam was started from collimators at $\mathrm{S}=3583$\,m (vertical halo) and $\mathrm{S}=3691$\,m (horizonal
    halo). The diagram follows the curvilinear coordinate frame of the machine and
    is integrated across all turns of all particles, i.e. the diagram wraps around.}
\end{figure*}

In this example, the collimation system is designed to work independently in the horizontal
and vertical planes. Therefore, each is simulated independently with the appropriate beam distribution
for each dimension. These can later be combined for total beam loss information.
Figure~\ref{fig:modellosses} shows the combined losses from the simulation of $2\times10^{5}$
protons with 100\,GeV energy in both the horizontal and vertical planes. By default,
the user must choose the physics list or sets of processes they require (none were
used for the optics validation as none were required). In this case, the Geant4
\emph{reference} physics list {``FTFP\_BERT''} was used that is a standard high
energy physics list provided by Geant4 that BDSIM provides an interface to. This provides
a wide variety of physics processes including electromagnetic, hadronic elastic, inelastic
and diffractive processes. In BDSIM, it is accessed using the following command:

\begin{lstlisting}
option, physicsList="g4FTFP_BERT";
\end{lstlisting}

Practically, the simulation was performed on the 500-core Royal Holloway Cluster in 2000
jobs that produced a total 60\,GB of ROOT format output. These were analysed individually
using \texttt{rebdsim} with the same analysis configuration ASCII file. The resulting 2000
\texttt{rebdsim} histogram files were then combined using the provided
\texttt{rebdsimCombine} tool. It should be noted that such large computational resources
are not required and one can easily simulate millions of events on a personal
computer depending on the size of the model and the physics used.

Figure~\ref{fig:modellosses} shows the proton impact and loss locations. The impact location is
determined as the
first point where a physics process is invoked along the step of the particle and the loss
point is the end of the trajectory of the primary proton. This may be either due to an inelastic
collision and subsequent fragmentation, or simply absorption. Additionally, the energy deposition
in all material in the accelerator is shown. A clear maximum in losses and energy deposition can
be seen in proximity to the collimation section between 3500\,m and 4000\,m. However, a clear
decaying tail of energy deposition is seen throughout the subsequent arc (in the direction of
the beam). For the first 1\,km of the machine there are very few particle losses but considerable
energy deposition. Such energy deposition would not be shown from only tracking primary particles
and is a unique feature of BDSIM.

All histograms are made on a `per-event' basis and the histograms shown are the mean
of the sampled number of events. The uncertainty for each bin is the standard error on
the mean. This allows the statistical sample simulated to be scaled to a realistic rate
accurately. Here, one event is one particle simulated. Any realistic rate scaling would
take into account the chosen phase space for the input beam distribution that was used
for increased simulation efficiency. This per-event analysis is only possible because of the
event-by-event storage of the output format and this is crucial to calculate the correct
statistical uncertainty associated with each bin in the histogram.

\begin{figure}
  \centering
  \includegraphics[width=8.5cm]{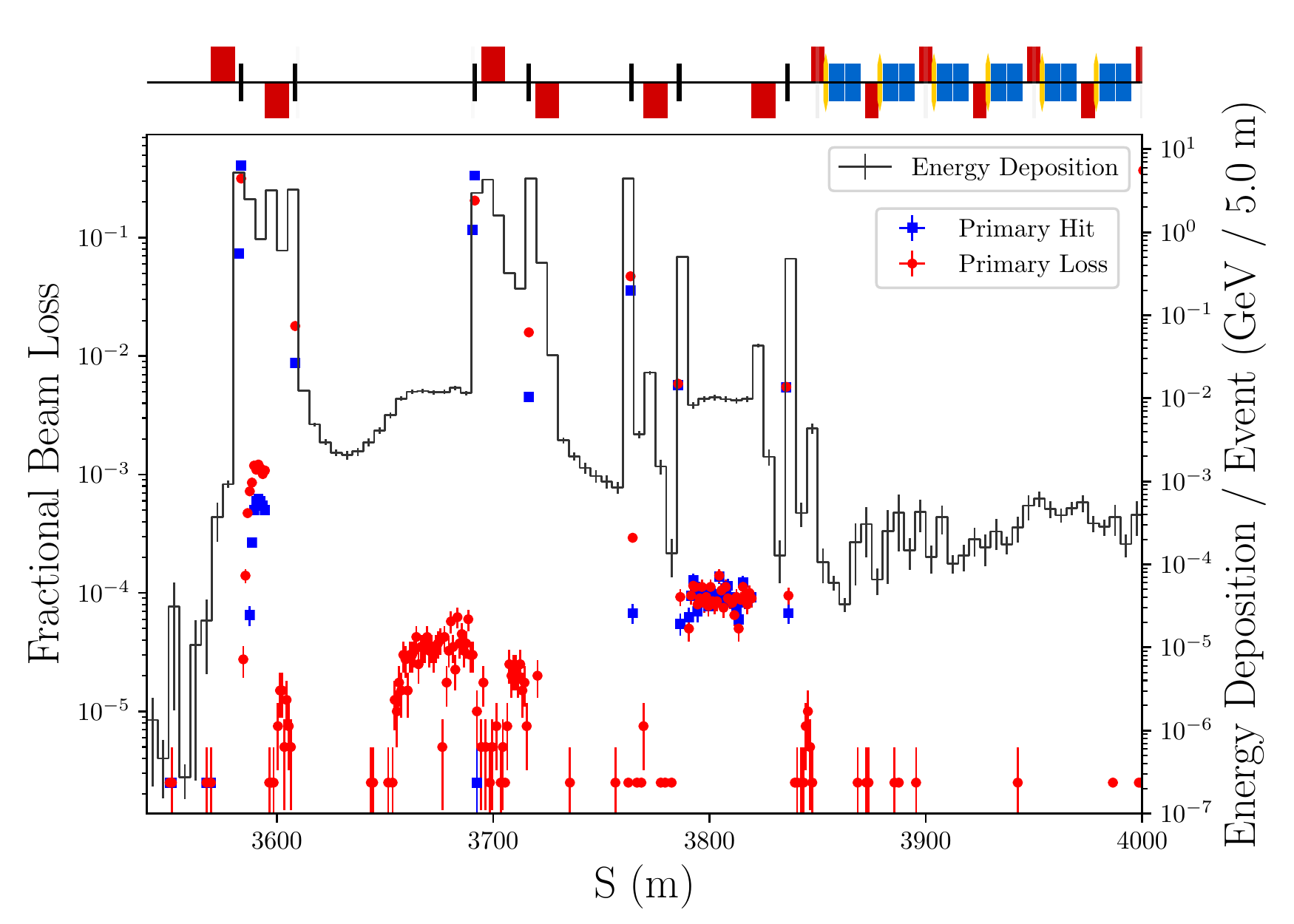}
  \caption{\label{fig:modellossescollimation}Collimation section detail showing
    combined beam loss from a horizontal and vertical halo simulation with
    $4\times10^{5}$ primary protons at 100\,GeV.}
\end{figure}

Looking at the collimation insertion region in more detail as shown in
Figure~\ref{fig:modellossescollimation}, the pattern of proton impacts
and losses can be seen more clearly. The `hits' and `losses' are quite different
indicating that primary protons can interact with a collimator and escape. Such
information could be used to improve the efficiency of the
simplistic collimation section in this model to better absorb the scattered
or leaked protons.

The information shown from the simulation is already a powerful
guide to the operating radiation produced in the machine. However, further
information is easily available that allows an even greater understanding
of the machine behaviour. By using the `sample' command on each collimator,
the complete distribution of particles after each element can be recorded.

\begin{figure}
  \centering
  \includegraphics[width=8cm]{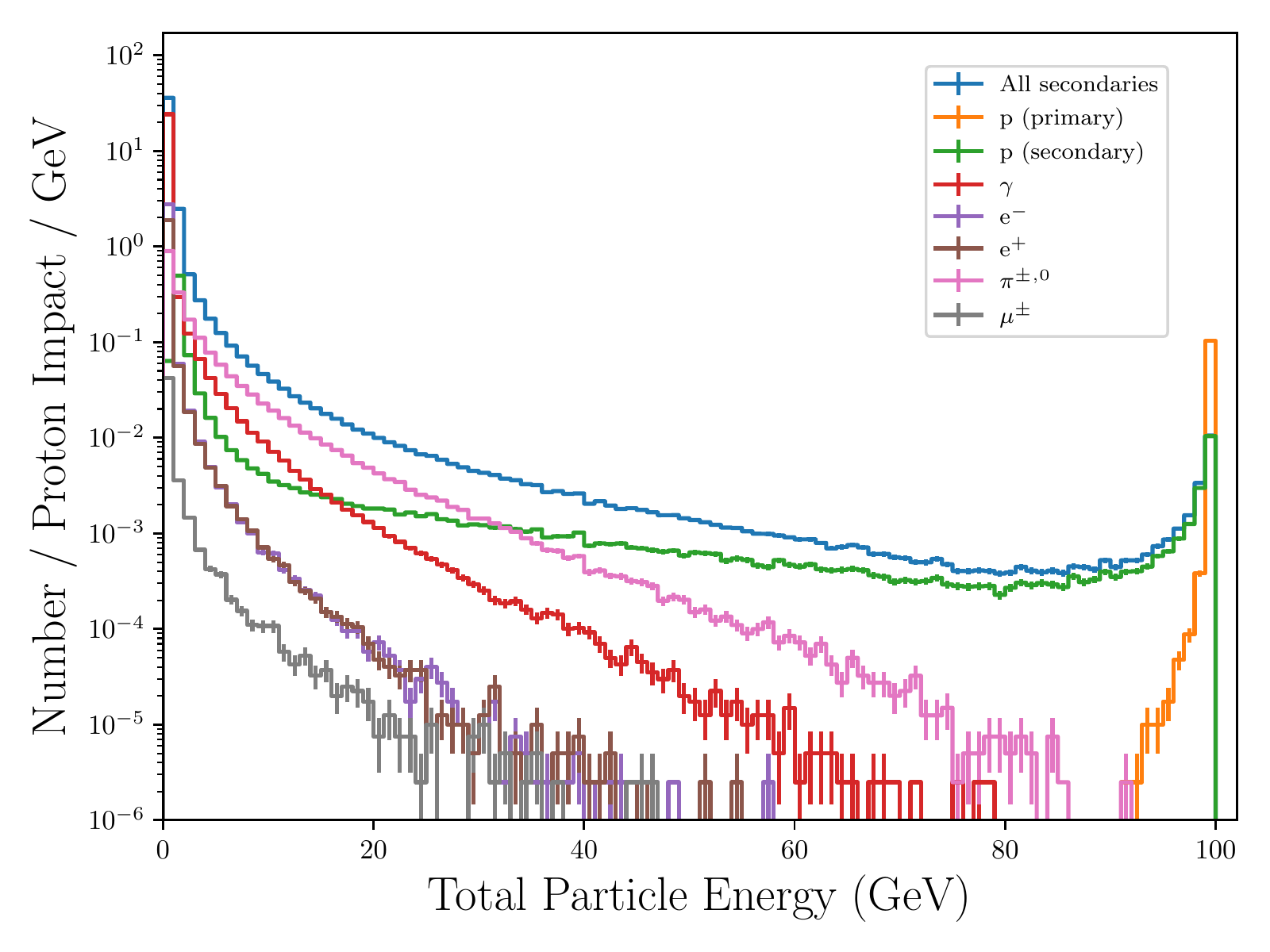}
  \caption{\label{fig:modelspectrum}Particle energy spectrum of all particles
    recorded after the primary horizontal collimator. The data is integrated for
    all 100 revolutions of the primary beam in the simulation.}
\end{figure}

Figure~\ref{fig:modelspectrum} shows the energy spectrum of particles recorded in a
sampler placed after the primary horizontal collimator for the horizontal halo
simulation. For each particle species a separate histogram was prepared with \texttt{rebdsim}
by defining a `selection' that acts as a filter. Here, the integer particle ID
from the Monte Carlo Numbering scheme~\cite{PDG:2018} was matched. The spectra
show significant fluxes of various high energy particles leaving the primary
collimator and in particular a much broader spectrum of primary protons than
were originally in the beam. These are of particular interest as they may
travel some distance before being lost in a position that would not be
immediately associated with the collimator in question.

\begin{figure}
  \centering
  \includegraphics[width=8cm]{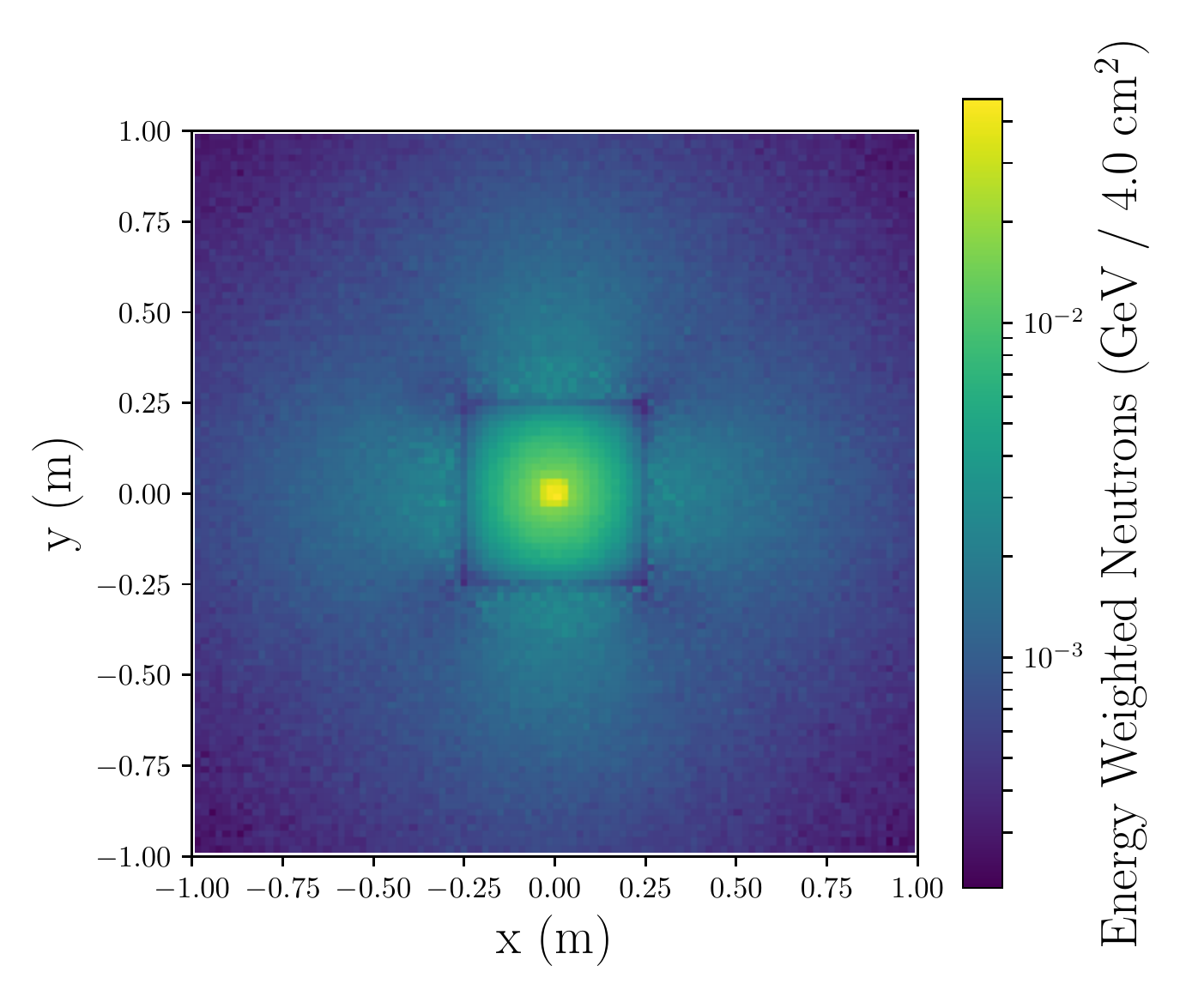}
  \caption{\label{fig:modelneutronflux}The energy-weighted neutron
    flux after the secondary horizontal collimator per proton impact
    on the primary collimator. Although the collimators have an
    asymmetric aperture, their outer shape is a square. Here the shadow
    of the square 0.5\,m wide collimators can clearly be seen.}
\end{figure}

Furthermore, for the purpose of damage and activation it is useful to look at the
neutron distribution as an example. Shown in Figure~\ref{fig:modelneutronflux} is
the 2D distribution of neutrons weighted by their total energy after the secondary
horizontal collimator from the horizontal halo simulation only. Here, a clear
shadow of the collimator can be seen. The \texttt{rebdsim} input syntax is shown below
to highlight the simplicity of making per-particle-species rate normalised
histograms from simulation data without the need for a complicated analysis.

\begin{lstlisting}
  Histogram2D
  Event.
  COLPRIMXEWFluxProtSecZoom
  {100,100}
  {-0.5:0.5,-0.5:0.5}
  COLPRIMX.y:COLPRIMX.x
  COLPRIMX.energy*(COLPRIMX.partID==2112&&COLPRIMX.parentID>0)
\end{lstlisting}

\begin{figure}
  \centering
  \includegraphics[width=8cm]{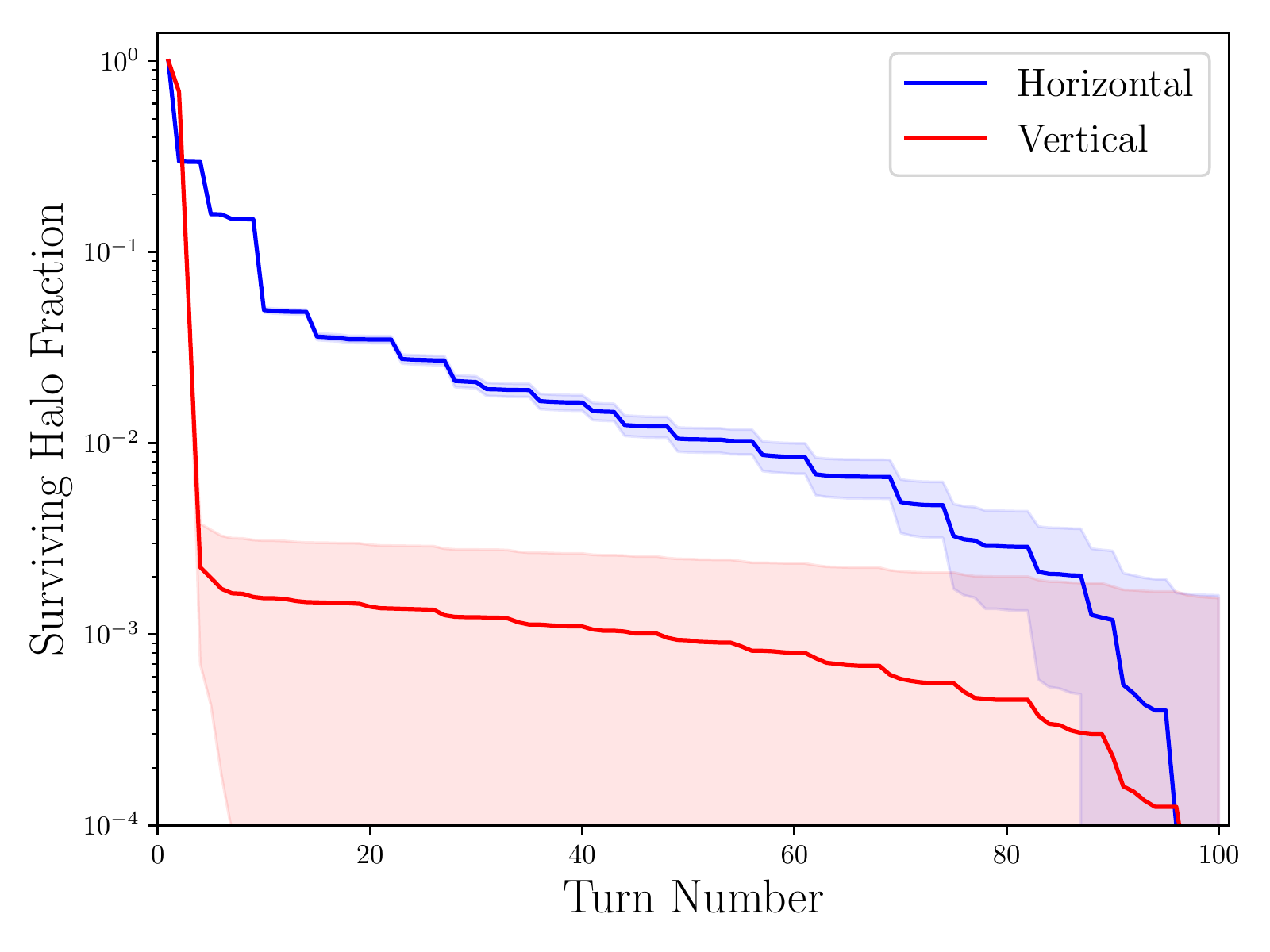}
  \caption{\label{fig:modelsurviving}Fraction of primary halo protons surviving
    per turn completed in the ring for both the horizontal and vertical
    halo simulations. The coloured band represents the statistical uncertainty
    from the simulation.}
\end{figure}

For circular models, the losses as a function of turn number are important.
BDSIM records the turn number associated with all data in the output for
circular models permitting turn-by-turn analysis. Figure~\ref{fig:modelsurviving}
shows the surviving fraction of the simulated halo beam after each turn. The
periodic steps are due to the tune of the circular machine and the figure
illustrates the different performance of the collimation system in each
plane. Should it be desired, the data presented in Figure~\ref{fig:modelspectrum}
and Figure~\ref{fig:modelneutronflux} can easily be filtered by turn number to
show the radiation on a particular turn or range of turns that may correspond
to a sharp increase of losses.

\section{Validation}
\label{sec:validation}

\subsection{Tracking Validation}
\label{ssec:trackingvalidation}

Tracking accuracy is crucial for physics
studies to correctly describe the machines being simulated. Imprecise tracking
in any single component would result in incorrect subsequent tracking in the machine.
This is particularly relevant to circular machines that would suffer
immensely from cumulative errors due to the high number of components and repeated
passage of particles.

Individual components in BDSIM as well as the optical functions of many large
current accelerators have been tested extensively in BDSIM over many years and
optical agreement is expected out-of-the-box. To demonstrate tracking validation,
the most obvious route is to compare with existing particle tracking software.
However, care must be taken to make such a comparison correctly and accurately.
In particular, most accelerator codes output data in ASCII format and the transfer of
numerical coordinates via such files can lead to truncation and severe
loss of precision. Furthermore, many tracking programs operate with an assumed
range of validity, such as the paraxial approximation, or can be used with different
integrators. \MADX{} PTC for example, can be used with the expanded or exact Hamiltonian,
which can lead to vastly different results. Several tracking programs have implicit
assumptions such as the primary particle being assumed to be fully relativistic.
Particle tracking codes often allow the user to choose between computational performance
and increased physical accuracy depending on their needs. With this knowledge in mind,
any comparisons between codes must be carefully performed with knowledge and
understanding of the limitations of particular software.

The validation of BDSIM tracking has been performed and ensures that:

\begin{enumerate}
\item The output of the implementation of an individual integrator matches
  that of the mathematical description as documented in the manual.
\item The automatic preparation and conversion of a model is correct and
  accounts for the often inconsistent output of the supported optical
  descriptions.
\item A converted model reproduces the expected optical functions throughout
  a lattice.
\end{enumerate}

In addition, to the caveats of using other tracking software, we must consider also
how information is recorded in BDSIM. In particle tracking software, an algorithm
is applied to a set of coordinates representing the traversal through an accelerator
component. The coordinates reported after this single traversal represent the particle
precisely at the end of the element. However, in BDSIM, we make a 3D model where
no 2D shapes can exist. It is not possible through the interfaces in Geant4 to
record particles at only one surface of a multi-faceted solid (e.g. one face of a cube).
We must also avoid coplanar faces or geometrical overlaps with a numerically resolvable
gap in-between each element for robust geometry navigation and tracking. In BDSIM,
the particle coordinates are recorded on the one step taken through the 1\,nm thin
sampler, which is placed after the element with a 1\,nm gap
between each surface. This difference in distance may introduce a very small
difference in coordinates and agreement is expected only to this degree. The distance
between surfaces in BDSIM was carefully chosen and the spacing of faces around
the end of each element in both the mass world and the several parallel worlds
chosen so as to avoid coplanar faces between worlds wherein Geant4 would incorrectly
navigate through the geometry.

Particle to particle comparisons show residuals at the level of $10^{-9}$ in
both position (m) and transverse (normalised) momentum for particles in the
paraxial approximation for the most common design of magnets (length and strength)
found in a large variety of accelerators. The comparison was made by specially
compiling BDSIM with double precision floating point output and increasing
the precision of the output from \MADX{} PTC.

Optical comparisons for a variety of accelerators including the LHC, the Accelerator
Test Facility 2 (ATF2) at KEK, Japan, the Diamond Light Source (DLS), UK,
show excellent agreement with both \MADX{} PTC and the optical functions from \MADX{}. The
LHC demonstrates accuracy through a large number of components. The ATF2 demonstrates
correct transport through a highly non-linear lattice where pole face rotations
and dipole fringe fields are important. The ATF2 magnifies a beam before creating
a nanometre-level focus, so any small deviations are readily observable. The
DLS demonstrates a machine whose performance
is highly dependent on chromatic and dispersive effects. When testing large lattices
it is common to witness a deviation in the mean in several parameters. This is due
to the finite sample of particles used, where the initial sample-based mean is non-zero.
Such a non-zero offset is then witnessed as it propagates. BDSIM includes an advanced
option where the sample deviation may be subtracted by pre-generating the full
distribution and calculating the sample mean from it. This is only valid for large
sample sizes and is used for developer optical comparisons.

\begin{figure*}
  \normalsize
  \centering
  \includegraphics[width=0.97\textwidth]{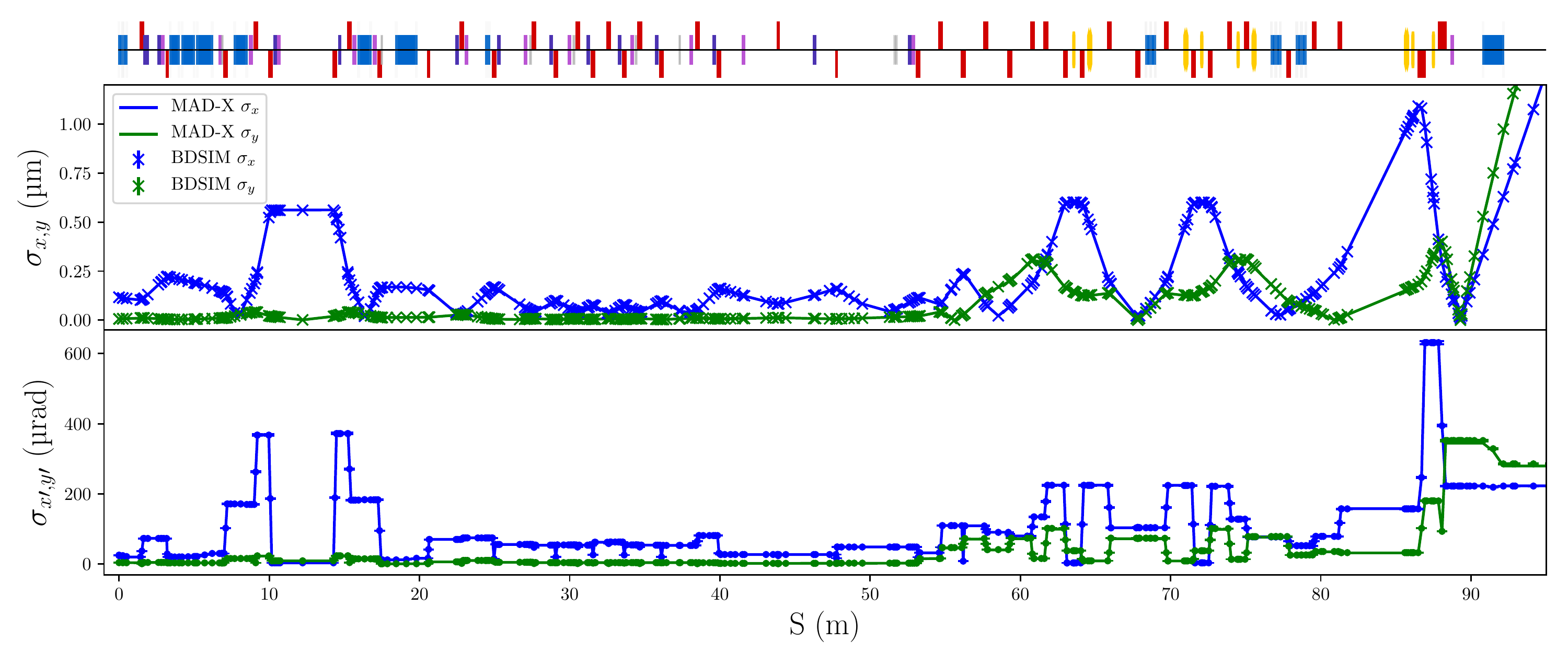}
  \caption{\label{fig:validationatf2} Comparison of beam size (top) and angular size
    (bottom) between tracking in BDSIM (points) and the \MADX{} model (solid line) throughout
    the Accelerator Test Facility 2 (ATF2) extraction line at KEK, Japan. After extraction
    from a damping ring, the beam progresses left to right along a single pass extraction line.
    The comparison shows excellent agreement throughout as the beam size is strongly manipulated.
    The effect of dipole pole face rotations is not visible directly but is evident in the agreement
  that would be coarsely wrong if this minuscule effect was omitted.}
\end{figure*}

Although the tracking routines provided are demonstrably accurate, as already
mentioned, the 3D model
must include a small numerically resolvable gap between each volume. For long term
tracking in circular accelerators, such geometrical effects accrue and can result
in inaccurate tracking. However, the purpose of BDSIM is not to perform long-term
tracking studies, but to simulate losses and subsequent radiation. As shown in
Section~\ref{sec:example}, a collimation simulation typically starts only with
particles that impact a collimator. For a typical LHC collimation study, up to
200 turns of tracking are permitted~\cite{Bruce:2014}. For a particle at approximately
one sigma of the nominal distribution, an offset of 3\,$\mathrm{\mu}$m is accrued after 200 turns
which is acceptable.

As tracking in a BDSIM model is ultimately performed in Cartesian coordinates, BDSIM
cannot be used for arbitrarily large circular machines. This is because the numbers for
each coordinate are stored as double precision floating point numbers. When a particle
makes an excursion to the furtherst part of the ring where there maybe an offset of tens
of kilometres, the $\sim$15 significant figures will lead to a lack of precision in
curvilinear coordinates. This is not foreseen as a problem as the largest existing
machine is the LHC, where the tracking has been shown to be acceptable for radiation
simulations.

A common concern in particle tracking software is whether the tracking is symplectic.
This is equivalent to precisely conserving energy and phase space volume after both a single
and multiple applications of a transfer map or algorithm. The integrators of the linear
elements in BDSIM are thick lens solutions and are symplectic. The higher order integrators
are low-order semi-implicit Euler integrators that are symplectic. These conserve energy
but cannot be used accurately for large step sizes and be physically accurate. In practice, the
numerical integration error is estimated and returned to Geant4 that limits the step size
in strongly varying magnetic fields. In validation testing, it was found that the accrued
geometrical errors dominate in larger circular models. In linear models, the tracking is
sufficiently accurate and symplecticity is not of concern.

In the case of circular machines, it is possible to use a one-turn-map with BDSIM. This is
an externally provided transfer map representing a single turn of the circular machine that
can be used to correct the particle coordinates on
each turn. If the particle completes a turn of the circular machine without any physics
process being invoked, the coordinates after the turn are calculated independently for that
turn using the map and the particle coordinates updated. This approach allows a maximum
error of approximately $1\,\mathrm{\mu}$m to be achieved for up to 10,000 turns of the LHC using a
14th order map calculated by \MADX{} PTC. It should be noted however, that if the lattice contains
non-linear elements, the map will typically be non-symplectic. However, with a high order map,
it would take a high number of turns outside the applicability of BDSIM to reach this limitation.

For the purpose of simulating accelerators smaller than the LHC and linear or single pass
sections, the tracking is highly accurate.

\subsection{Quality Assurance}
\label{ssec:qualityassurance}

To be used in a predictive capacity, aside from validation there must be assurance
that any changes or modifications to the software do not affect any known outcome.
Any new additions must also be demonstrated to work reliably with a variety of input.
BDSIM uses two main types of tests to ensure the quality and stability of the software,
namely unit and regression tests.

Unit tests are minimal input examples that exploit a particular feature. Running such
an example in BDSIM forces use of the feature in the BDSIM code and tests it executes
without problem. For every test of correct input there are matching tests of incorrect
input that BDSIM should reject. The condition for the test to pass in this case is that
BDSIM should exit. At the time of writing there are 689 unit tests that cover 87.61\% of
the C++ code in BDSIM. The parts not covered are identified as parts that are for external
interfaces and are tested elsewhere or for interactive visualisation that cannot be tested
automatically, but are however covered by regular use. The tests are defined in CMake~\cite{CMAKE-website}
as part of the build system throughout the examples directory and are run with CTest, which
is part of CMake. Every example also forms
a unit test. The majority of tests are run single-threaded within 1hr and are automatically
run at the Royal Holloway Faraday Cluster up to every hour if the code is updated in the
public-facing git repository. The results are reported to a CDash website hosted at Royal
Holloway, where code coverage and test results and output can be inspected. Developers are
automatically notified of newly failing tests due to commits they have made. Several
builds for Scientific Linux 6 and 7 and Ubuntu operating systems are made. In addition,
separate builds for Geant 4.10.1 through to 4.10.5 are made and all tests running for each.
This variety gives very good coverage of operating systems and environments that are expected
to be used for BDSIM.

In addition to unit tests, regression test are used. Unit tests ensure that there is no
syntactic or logical problem with the code, but not that it achieves the correct result.
Regression tests are used to compare the output of a fresh simulation with a manually
validated result. To make the comparison, a special tool, the \texttt{comparator}, is
included with BDSIM. This is capable of dynamically inspecting all the data structures
found in a BDSIM output file and comparing them.
In the case of coordinates at a sampler plane, individual entries can
be compared within the numerical precision of the output. In these cases, tests are conducted
with the same random number generator seed, so the output should be identical. In the case
where two histograms are compared, the contents can be compared statistically using the
Student's T-Test.

Several reference output files are created for large accelerators such as the ATF2,
DLS and the LHC included in the examples. These contain a large variety of elements
with a large variety of settings, so provide a high level of coverage over tracking
algorithms. Any change in these will be shown by the single-turn or pass coordinates
being different. The developers can then investigate to identify the precise cause of
the discrepancy. Aside from complete lattices, many references files are included for
key features of BDSIM as well as several examples that use a combination of many options.

Aside from developer-side testing, BDSIM uses a publicly visible software repository
using the git software versioning software~\cite{GIT-website}. The BDSIM website,
(\url{http://bitbucket.org/jairhul/bdsim}),
also hosts a public issue tracker where anyone can report an issue and is regularly
updated by the developers. The use of the git software versioning system in a public
repository allows users to review and understand both how BDSIM works as well as
possibly contribute. The authors have found the open-source nature to be invaluable
in fostering a community that understands and contributes to the code.

Usage of the code is documented in an included manual prepared using the commonly used
Sphinx documentation system~\cite{Sphinx-website}. This allows export of the manual as
both an HTML website and \LaTeX{} document that is rendered as a PDF, ensuring both are
consistent. The classes and structure of the code is documented using
Doxygen~\cite{Doxygen-website} that parses comments in the source files. The code is
fully documented using this system and the output Doxygen website is publicly hosted.

\section{Conclusions}
\label{sec:conclusions}

BDSIM is an open source C++ code that makes use of widely used and current
high energy physics software to provide a mixed 3D and accelerator tracking
simulation suitable for predicting the radiation in and around an accelerator.

It provides a unique capability to accurately simulate radiation transport in
an accelerator environment. It significantly simplifies the construction of a
3D Geant4 model by providing a
library of generic but scalable accelerator components as well as the appropriate
fields and numerical integrators for the most accurate particle tracking. Additional
thin elements such as multipoles and magnet fringe elements ensure accurate
magnetic tracking that matches the most commonly used accelerator tracking
codes. Furthermore, the ability to track all particles through the accelerator
with arbitrary step sizes is unique and allows prediction of the radiation
transmitted through and around an accelerator. The data format and provided
analysis tools are built on ROOT, software used commonly in particle physics,
that is highly scalable and suitable for long term data storage and very large
data quantities.

The included Python utilities allow preparation of models in minutes that would
normally take several weeks to prepare manually and are easily extendable
with further user input and customisation. As BDSIM as well as all of its
dependencies are open source, the program is
highly transparent to the user with complete and regularly updated documentation.
Furthermore, the software can be easily extended with user's C++ code.

BDSIM has been engineered to be applicable to both the lowest and highest energy
accelerators in use today. It provides a unique ability to perform a mixed
accelerator and radiation transport simulation as well as remove many obstacles
for developing a  beam line model. The software was written in a way to support
the widest set of simulations and beam lines.

\section{Acknowledgements}

The development of BDSIM has received funding from the John Adams Institute at
Royal Holloway. This work was supported by Science and Technology Research
council grant ``The John Adams Institute for Accelerator Science'' ST/P00203X/1
and Impact Acceleration Account.  It was also supported in part by EU FP7 EuCARD-2, Grant
Agreement 312453 (WP13, ANAC2) and STFC, United Kingdom.
Thank you to J.~Albrecht, S.~E.~Alden, D.~Brewer, A.~Cooper, W.~Parker, J.~Van~Oers,
M. Porter and S.~Williams for their contributions and testing.






\bibliographystyle{elsarticle-num}
\bibliography{bdsim}







\end{document}